\pdfoutput=1

\documentclass{aa}

\usepackage{graphicx}
\usepackage[varg]{txfonts}
\usepackage{amsmath}
\usepackage{verbatim}
\usepackage{latexsym}
\usepackage{multirow}
\usepackage{fixltx2e}
\usepackage{color}
\usepackage{rotating}

\newcommand{\jurcak}{{Jur{\v c}{\'a}k}}

\begin{document}

\title{No universal connection between the vertical magnetic field and the umbra-penumbra boundary in sunspots}

\author{B. L\"optien\inst{1}
\and A. Lagg\inst{1}
\and M. van Noort\inst{1}
\and S.~K. Solanki\inst{1,2}}

\institute{Max-Planck-Institut f\"ur Sonnensystemforschung, Justus-von-Liebig-Weg 3, 37077 G\"ottingen, Germany
\and School of Space Research, Kyung Hee University, Yongin, Gyeonggi, 446-701, Republic of Korea}

\date{Received <date> /
Accepted <date>}

\abstract {It has been reported that the boundary between the umbra and the penumbra of sunspots occurs at a canonical value of the strength of the vertical magnetic field, independently of the size of the spot. This critical field strength is interpreted as to be the threshold for the onset of magnetoconvection.}
{Here we investigate the reasons why this criterion, also called the \jurcak{} criterion in the literature, does not always identify the boundary between umbra and penumbra.}
{We perform a statistical analysis of 23 sunspots observed with Hinode/SOT. We compare the properties of the continuum intensity and the vertical magnetic field between filaments and spines and how they vary between spots of different sizes.}
{We find that the inner boundary of the penumbra is not related to a universal value of the vertical magnetic field. The properties of spines and filaments vary between spots of different sizes. Both components are darker in larger spots and the spines exhibit stronger vertical magnetic field. These variations of the properties of filaments and spines with spot size are also the reason for the reported invariance of the averaged vertical magnetic field at 50\% of the mean continuum intensity.}
{The formation of filaments and the onset of magnetoconvection are not related to a canonical value of the strength of the vertical magnetic field. Such a seemingly unique magnetic field strength is rather an effect of the filling factor of spines and penumbral filaments.}

\maketitle

\section{Introduction}

The penumbrae of sunspots exhibit a complex structure. They consist of bright filaments with a horizontal magnetic field, interlaced with dark spines with a more vertical magnetic field \citep{1993ApJ...403..780T,1993A&A...275..283S,1993ApJ...418..928L,2013A&A...557A..25T}. While it is widely accepted that the penumbral filaments are a manifestation of overturning magnetoconvection \citep{2008ApJ...677L.149S,2008A&A...488L..17Z,2009ApJ...691..640R,2011ApJ...729....5R,2014ApJ...785...90R}, the mechanisms that are responsible for the formation of the penumbra are not well understood yet.

\citet{1995MNRAS.273..491R} modeled the transition from pores to sunspots as a bifurcation process, with the penumbra forming if the inclination of the outer edge of the flux tube exceeds a certain threshold. The inclined field lines might then be pumped downwards by convection \citep{2002Natur.420..390T,2004ApJ...600.1073W,2008ApJ...686.1454B}. Based on Hinode observations, \citet{2014PASJ...66S..11K} could distinguish between three different scenarios leading to the formation of a penumbra, active accumulation of magnetic flux \citep[the pore accumulates more magnetic flux, which leads to a larger inclination at the outer boundary of the flux tube, this was also suggested by][]{1995MNRAS.273..491R,1998ApJ...507..454L,2010A&A...512L...1S}, rapid emergence of magnetic fields (having the same polarity as the pore), or the appearance of twisted or rotating magnetic flux tubes (leading to a strongly twisted penumbra).

There are strong indications from observations that magnetic fields in the chromosphere play a major role in the formation of the penumbra. An overlying magnetic field in the chromosphere could trap newly emerging flux in the photosphere and prevent it from becoming vertical \citep{1998ApJ...507..454L,2013ApJ...769L..18L}. Alternatively, a magnetic canopy in the chromosphere preceding the formation of the penumbra might sink down later and form the penumbra \citep{2012ApJ...747L..18S,2013ApJ...771L...3R,2014ApJ...784...10R,2016ApJ...825...75M}. A strong influence of the chromosphere on penumbra formation is further supported by numerical simulations, where the extent of the penumbra strongly depends on the details of the top boundary condition \citep{2012ApJ...750...62R,2020arXiv200403940J}.

During the decay of the penumbra, the magnetic field in the penumbra becomes more vertical \citep{2014ApJ...796...77W,2018A&A...614A...2V,2018A&A...620A.191B} and it might rise up to the chromosphere \citep{2008ApJ...676..698B}.

One important aspect that has to be explained by any successful model of the penumbra is the sharp transition from the umbra to the penumbra, particularly between the umbra and bright penumbral filaments. Recently, it was claimed that the umbra-penumbra (UP) boundary is connected to a fixed value of the strength of the vertical magnetic field component $B_{\rm z}$. \citet{2011A&A...531A.118J} reported that $B_{\rm z}$ varies significantly less along the UP boundary than the total magnetic field strength. The average vertical magnetic field seems to assume a canonical value of $B_{\rm thr} = 1867$~G at the UP boundary (defined as where $I_{\rm C} = 0.5$) for all spots, with no indications of a dependence on the size of the spots \citep{2011A&A...531A.118J,2018A&A...611L...4J,2018A&A...620A.104S,2020arXiv200409956L}.

This so-called \jurcak{} criterion is interpreted to be the threshold for the onset of magnetoconvection. According to this, magnetic flux tubes with a weaker vertical field than $B_{\rm thr}$ become unstable and bend over to form filaments, in agreement with the fallen flux tube model suggested by \citet{1992ApJ...388..211W}. In addition, \citet{2019ApJ...873L..10M} used the stability criterion of \citet{1966MNRAS.133...85G} to show that a vertical magnetic field of $B_{\rm thr} = 1867$~G is indeed sufficient to suppress convection within sunspot umbrae. The \jurcak{} criterion is very interesting, as it could be a simple and straightforward method for separating between umbra and penumbra.

However, parts of the umbra of some sunspots have a weaker $B_{\rm z}$ than allowed by the \jurcak{} criterion. These regions are interpreted by \citet{2018A&A...611L...4J} to be unstable against magnetoconvection and about to be transformed into penumbra. Indeed, there are some observational indications for such a behavior, such as the formation of penumbra in an emerging sunspot \citep{2015A&A...580L...1J}, or a pore with a weak magnetic field that gets transformed into orphan penumbra \citep{2017A&A...597A..60J}. Similarly, Hinode observations of a decaying sunspot show that the vertical field at the UP boundary is always weaker than $B_{\rm thr}$ during the decay phase \citep{2018A&A...620A.191B}. However, a definite proof for a connection between the \jurcak{} criterion and the stability of umbrae is still missing.

In this study, we revisit the question whether the \jurcak{} criterion is a good criterion for identifying the UP boundary. Using a sample of 23 individual sunspots observed by Hinode, we show that there is no unique value of $B_{\rm z}$ which outlines the UP boundary for all spots in a consistent way. As explained above, parts of the umbra with a lower $B_{\rm z}$ than given by the \jurcak{} criterion are interpreted to be about to be converted into penumbra. We present a case study of the decaying sunspot AR~10953, which might challenge this interpretation.

In the second part of this paper, we show that the observed constant value of the average $B_{\rm z}$ at $I_{\rm C} = 0.5$ for spots of different sizes is not related to the onset of magnetoconvection. Instead, it is caused by differences in the brightness of penumbral filaments between spots of different sizes, with large spots harboring darker penumbral filaments.

\section{Data}
\begin{table*}
\caption{Overview of the sunspot observations used in this study.}
\label{tab:spots}
\centering
\begin{tabular}{l l c c c}
\hline\hline
NOAA & Date & $\theta$& umbral area & evolutionary state \\
     &      & [deg]   & [Mm$^2$]    & \\
\hline
10921  & 2006.11.05 & 22 &  54 & stable \\
10923  & 2006.11.12 & 31 & 541 & stable \\
       & 2006.11.13 & 15 & 550 & stable \\
       & 2006.11.14 &  8 & 571 & stable \\
       & 2006.11.15 & 13 & 587 & stable \\
       & 2006.11.16 & 30 & 558 & stable \\
10926  & 2006.12.03 & 24 &  33 & forming \\
10930  & 2006.12.13 & 29 & 366 & stable \\
       & 2006.12.14 & 41 & 350 & stable \\
10933  & 2007.01.04 & 16 & 155 & stable \\
       & 2007.01.06 &  9 & 159 & stable \\
       & 2007.01.07 & 24 & 149 & stable \\
10944  & 2007.02.28 &  2 &  62 & stable \\
       & 2007.03.01 & 11 &  64 & stable \\
       & 2007.03.02 & 16 &  64 & stable \\ 
       & 2007.03.03 & 44 &  55 & stable \\
10953  & 2007.04.30 & 13 & 290 & stable \\
       & 2007.05.01 &  7 & 284 & stable \\
       & 2007.05.02 & 12 & 232 & stable \\
       & 2007.05.03 & 24 & 212 & stable \\
       & 2007.05.04 & 37 & 195 & stable \\
10960  & 2007.06.10 & 34 &  44 & stable \\
       & 2007.06.11 & 54 &  36 & stable \\
       & 2007.06.12 & 57 &  34 & stable \\
10969  & 2007.08.27 & 12 &  66 & decaying \\
       & 2007.08.28 & 16 &  57 & decaying \\
11039a & 2009.12.27 & 44 &  46 & forming \\
       & 2009.12.28 & 34 &  39 & forming \\
       & 2010.01.01 & 29 &  55 & forming \\
11039b & 2010.01.01 & 29 &  49 & forming \\
11041  & 2010.01.26 & 20 &  35 & decaying \\
11106  & 2010.09.16 & 27 &  34 & decaying \\
11113  & 2010.10.20 & 14 &  70 & stable \\
       & 2010.10.22 & 35 &  47 & stable \\
11117a & 2010.10.25 & 18 &  44 & stable \\
       & 2010.10.27 & 25 &  33 & stable \\
       & 2010.10.28 & 35 &  33 & stable \\
11117b & 2010.10.27 & 25 &  99 & forming \\
       & 2010.10.28 & 35 &  88 & forming \\
11131  & 2010.12.10 & 41 & 268 & stable \\
11195  & 2011.04.26 & 26 & 103 & stable \\
11279  & 2011.08.31 &  6 &  46 & decaying \\
11360  & 2011.11.28 & 21 &  41 & decaying \\
11361  & 2011.12.03 & 34 &  41 & stable \\
11363  & 2011.12.06 & 25 & 190 & stable \\
11419  & 2012.02.18 & 37 &  23 & decaying \\
11536  & 2012.07.31 & 34 &  28 & forming \\
11560  & 2012.09.02 &  9 &  69 & forming/decaying \\
\hline
\end{tabular}
\end{table*}

We perform a statistical analysis using 48 observations of 23 individual sunspots that were performed in normal mode with the spectropolarimeter on the Solar Optical Telescope \citep[SOT/SP,][]{2007SoPh..243....3K,2008SoPh..249..167T,2008SoPh..249..233I,2013SoPh..283..579L} onboard the Hinode spacecraft between 2006 and 2012. This instrument conducts spectropolarimetric observations using the Fe~I line pair at $6301.5$~\AA \ and $6302.5$~\AA. Table~\ref{tab:spots} lists the sample spots that were analyzed. In the table, the first column lists the NOAA active region number, the second one indicates the date when the spot was observed, the third one shows the heliocentric angle $\theta$, and the last column gives the total area of the umbra $A_{\rm U}$ in the spot (we define an area to be umbra if the continuum intensity is lower than 50\% of the mean continuum intensity in the quiet Sun). In some cases, multiple spots within an active region were observed. We distinguish between the individual sunspots by adding a letter (a or b) to the NOAA number. We manually characterize the evolutionary state of the sunspots as being stable, decaying, or forming using continuum intensity images from helioviewer~\footnote{https://helioviewer.org/}. Spots are defined as being stable if they do not form or decay over the course of the disk passage, when they were observed.

We inverted the observed Stokes parameters to derive the height-dependent full magnetic field vector. We used the spatially coupled version of the SPINOR code \citep{2000A&A...358.1109F,2012A&A...548A...5V,2013A&A...557A..24V} under the assumption of local thermodynamic equilibrium (LTE), with three nodes in optical depth for all atmospheric parameters, placed at $\log{\tau} = -2.5,-0.9,0$  \citep[cf. ][]{2013A&A...557A..25T}. We also applied a Lucy–Richardson deconvolution to the continuum intensity images in order to achieve a consistent spatial resolution between the maps of the intensity and the ones of the inverted atmospheric parameters. We resolved the $180^\circ$ azimuthal ambiguity as described in \citet{2018A&A...619A..42L} by using the Non-Potential Magnetic Field Computation method \citep[NPFC,][]{2005ApJ...629L..69G} and transformed the magnetic field vector to the local reference frame.

Some of the larger spots in our sample exhibit molecular lines and a high noise level in parts of the umbra. In these regions, the inversion is not reliable. Since these regions are not in the close vicinity of the UP boundary, this issue does not affect the results in this paper.

Our analysis is slightly different to the one of \citet{2018A&A...611L...4J}. Both studies are based on Hinode SOT/SP observations. However, there are differences in the inversion. \citet{2018A&A...611L...4J} performed their inversions using the SIR code \citep[Stokes inversion based on response functions, ][]{1992ApJ...398..375R} and allowed only the inverted temperature to vary with height. All other parameters (including the magnetic field vector) were assumed to be independent of height.

\section{Identifying the boundary of the umbra}
\subsection{Dependence of the umbra-penumbra boundary on the spot size}\label{sect:boundary}

\begin{figure*}
\centering
\includegraphics[width=17cm]{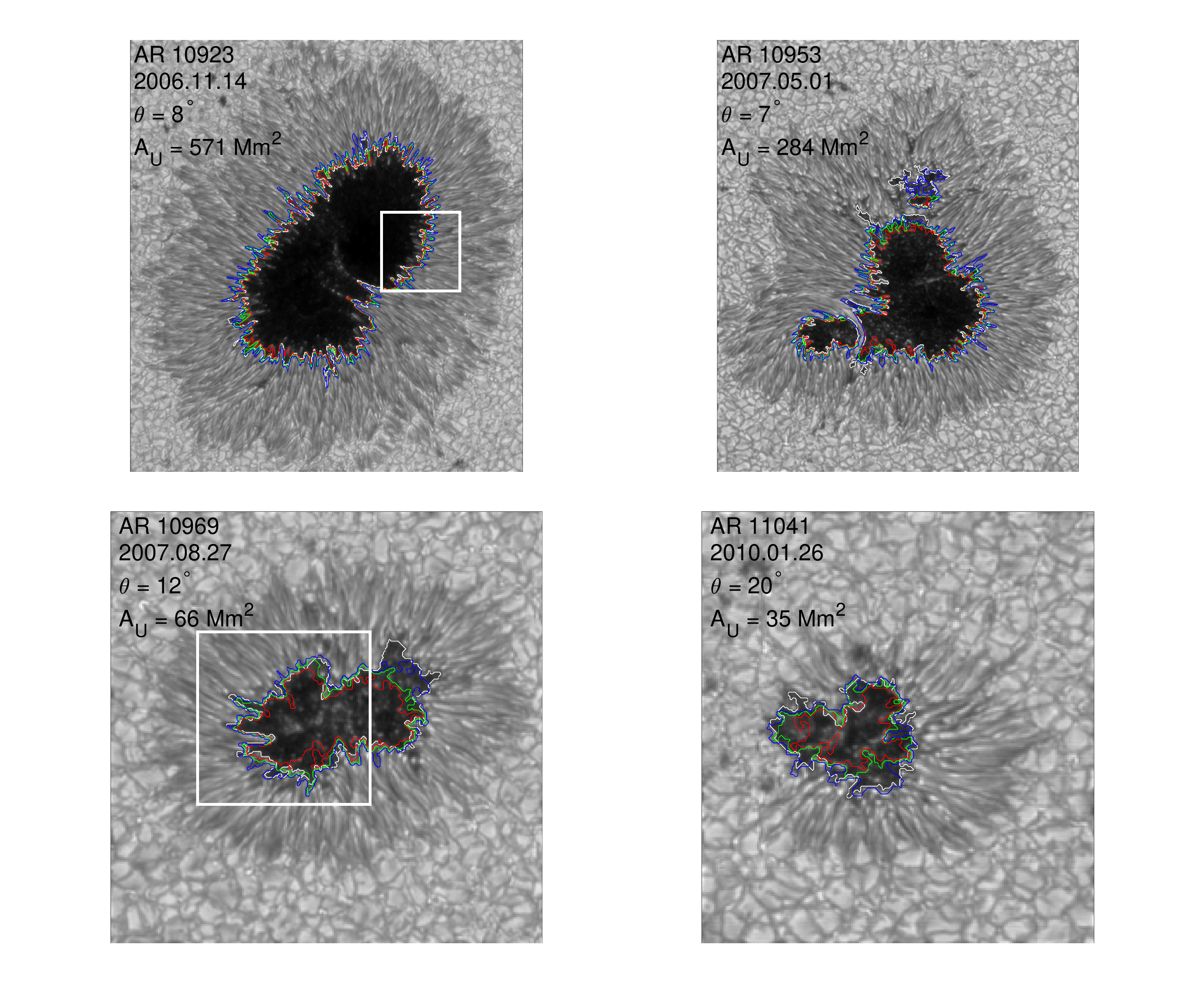}
\caption{Maps of the continuum intensity of selected sunspots. The white contours indicate a continuum intensity of 50 \% of the quiet Sun level. The other contours correspond to different strengths of the vertical magnetic field, evaluated at $\log \tau = -0.9$. Blue: 1650~G, green: 1750~G, and red: 1867~G. The white rectangles highlight regions of AR~10923 and AR~19069 that are shown in more detail later in the paper.}
\label{fig:spots}
\end{figure*}

\begin{figure}
\centering
\resizebox{\hsize}{!}{\includegraphics{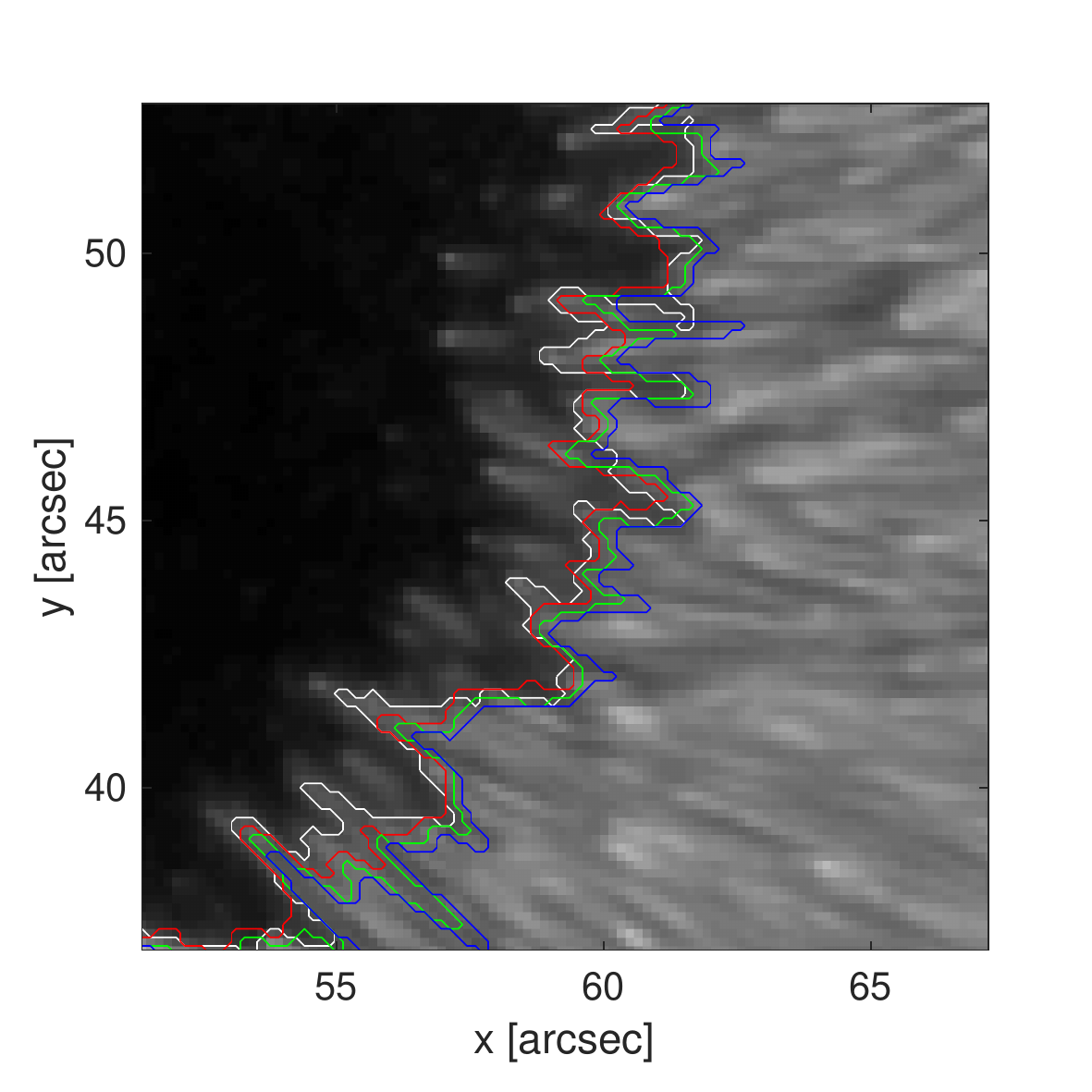}}
\caption{Zoom into a region of AR~10923, indicated by the white rectangle in the top left panel in Figure~\ref{fig:spots}. The contours are defined in the same way as in Figure~\ref{fig:spots}.}
\label{fig:10923_zoom}
\end{figure}

In this section, we evaluate the performance of a fixed threshold of $B_{\rm z}$ for defining the UP boundary. \citet{2018A&A...611L...4J} inferred a constant threshold of the strength of the vertical magnetic field of $B_{\rm thr} = 1867$~G (using their inversions, where $B_{\rm z}$ is independent of height) at the umbral boundary (which was defined as where $I_{\rm C} = 0.5$ in the wavelength range covered by Hinode SOT/SP). We apply the \jurcak{} criterion to the magnetic field at an optical depth of $\log \tau = -0.9$, which is the node of the inversion that is the most sensitive to the magnetic field (see also Section~\ref{sect:tau}). The maps of $B_{\rm z}$ exhibit noise on the scale of individual pixels, which affects the identification of the UP boundary. Hence, we reduce the noise before defining the umbral boundaries by convolving the maps of $B_{\rm z}$ with a Gaussian with $\sigma = 1$~pixels.

Figure \ref{fig:spots} shows maps of the continuum intensity of selected spots that were analyzed in this study with contours indicating the UP boundary. For comparison, we also show in white the UP boundary that was defined using a threshold of 50\% of the continuum intensity of the quiet Sun, For completeness, the maps for all spots are shown in Appendix~\ref{sect:maps}. In most cases, a threshold of 1867~G for $B_{\rm z}$ (as indicated by the red contours) is a good method for identifying the UP boundary. However, in some cases, the contour defined using the \jurcak{} criterion lies within the umbra, particularly for small spots (see,  e.g, AR~10969 or AR~11041 in Figure~\ref{fig:spots}). This indicates that $B_{\rm thr}=1867$~G is too large for some spots, as already reported by \citet{2018A&A...611L...4J}.

The threshold of $B_{\rm thr} = 1867$~G might not apply for our data, though, because of the differences in the inversion method ($B_{\rm z}$ changes with optical depth in our inversion). In the following, we test whether a lower threshold for $B_{\rm z}$ is more successful in identifying the UP boundary. The green and blue contours in Figure~\ref{fig:spots} are defined for $B_{\rm z} = 1750$~G and 1650~G, respectively. Indeed, a lower threshold for $B_{\rm z}$ works better for smaller spots, but even the lowest value that we tested (1650~G) is in some cases too high for identifying the UP boundary (e. g., AR~10969 in Figure~\ref{fig:spots}). In addition, this threshold is already too low for some of the larger spots. The contours defined using 1750~G or 1650~G extend far into the penumbra for large spots, assigning a large fraction of the spines of these spots to be part of the umbra (e. g., AR~10923, see top left panel of Figure~\ref{fig:spots} or Figure~\ref{fig:10923_zoom}).

Figure~\ref{fig:spots_all} suggests that the strength of the vertical magnetic field at the UP boundary depends predominantly on the size of the sunspot. We do not see any indications that the shape of the contours of constant $B_{\rm z}$ is affected by the evolutionary state of the sunspots. This is in contradiction to the hypothesis of \citet{2018A&A...611L...4J}, according to which a failure of the \jurcak{} criterion is connected to spot decay.

The observed dependence of the shape of the contours for a fixed $B_{\rm z}$ on spot size is not affected by the distance from disk center. There are differences between small and large spots, both close to disk center (e. g., AR~10969 and AR~10923 in Figure~\ref{fig:spots}) and closer to the limb (e. g., AR~10960 and AR~10930 in Figure~\ref{fig:spots_all}). However, in sunspots closer to the limb, the contours for a constant $B_{\rm z}$ are slightly shifted towards disk center (see, e. g., AR~10960 in Figure~\ref{fig:spots_all}). This effect was already noticed by \citet{2018A&A...611L...4J} and is probably caused by projection effects and by the increased formation height of the Fe~I lines when observing away from disk center.

Similarly, there are systematic differences between small and large spots when using the continuum intensity for defining the UP boundary. As can be seen in Figure~\ref{fig:10923_zoom}, a threshold of $I_{\rm C}=0.5$ assigns the innermost parts of some penumbral filaments of large spots to be part of the umbra. This does not occur in small spots like AR~11041 (see bottom right panel of Figure~\ref{fig:spots}). These dark parts of the penumbral filaments in large spots can have continuum intensities as low as $0.4$, which is comparable to the continuum of the umbra of some small spots. This suggests that there is no single value for the continuum intensity that can be used to identify the UP boundary for small and large spots in a consistent way.

\section{Intensity and vertical magnetic field at the UP boundary} \label{sect:hist}
\subsection{Brightness of penumbral filaments}
One of the main reasons why the \jurcak{} criterion is believed to be suitable for identifying the UP boundary is the observation by \citet{2018A&A...611L...4J} that the average $B_{\rm z}$ at $I_{\rm C} = 0.5$ is an invariant value and does not depend on the size of the umbral cores. This value of $B_{\rm z}$ is interpreted to be a threshold for the onset of magnetoconvection, which leads to the conversion of penumbra into umbra. However, this hypothesis has not been proven yet. As shown in the previous section, we do not see any indications of a connection between a failure of the \jurcak{} criterion in outlining the UP boundary and spot decay. In the following, we present an alternative explanation for the observed invariance of $B_{\rm z}$, which is based on the variation of the continuum intensity and of the strength of the vertical magnetic field between spots of different sizes. Hence, we start our discussion by evaluating how these parameters change between spots of different sizes, particularly for the penumbral filaments.

Large spots are generally darker than smaller ones \citep{1992sers.conf..130B,1993SoPh..146...49B,1994ApJ...432..403C}. This is particularly the case for the umbra, where the strength of the magnetic field is also higher for larger spots \citep[e. g.][]{1992SoPh..141..253K,2002SoPh..207...41L,2012A&A...541A..60R,2014SoPh..289.1477S,2014ApJ...787...22W,2014A&A...565A..52K,2015A&A...578A..43R}. In addition, \citet{2007A&A...465..291M} found the brightness of the penumbra to decrease with increasing spot size for spots with a radius larger than $10''$. In order to understand the invariance of the mean $B_{\rm z}$, we need to investigate how the properties of the penumbral filaments depend on the size of the spot. Figures~\ref{fig:spots} and~\ref{fig:10923_zoom} already indicate that penumbral filaments in large spots can be darker than in smaller ones. The continuum intensity of some parts of the penumbral filaments of AR~10923, for example, is lower than the threshold of $I_{\rm C} = 0.5$ that was used for defining the UP boundary. This is not the case in small spots, such as AR~11041 in Figure~\ref{fig:spots}.

\begin{figure}
\centering
\resizebox{\hsize}{!}{\includegraphics{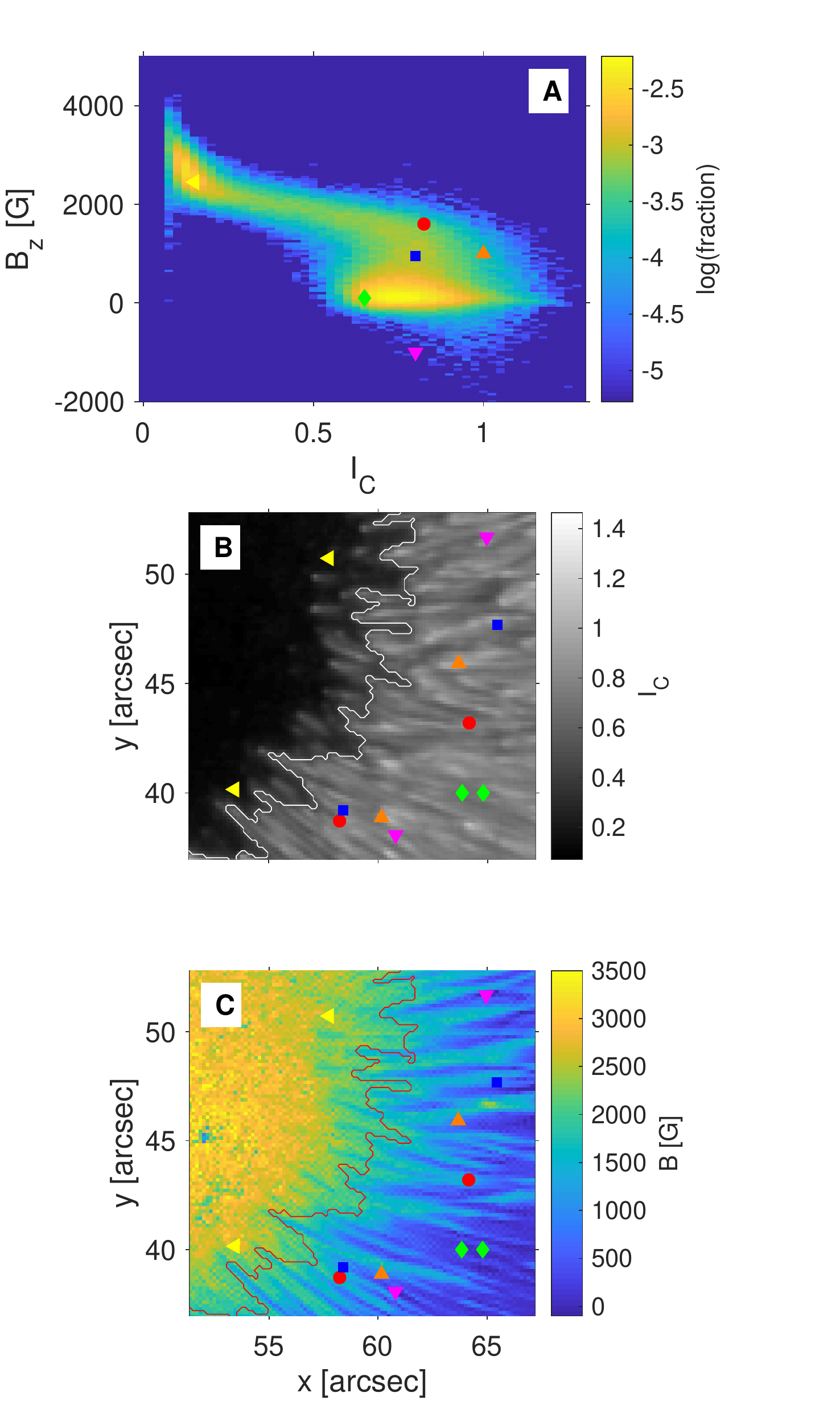}}
\caption{{\it Panel~A:} 2D histogram of the vertical magnetic field (at $\log \tau -0.9$) and the continuum intensity of AR~10923 observed on 14 November 2006 plotted on a logarithmic scale. Also shown are maps of the continuum intensity ({\it panel~B}) and of $B_{\rm z}$ at $\log \tau = -0.9$ ({\it panel~C}) of the spot. The field-of-view of these maps is the same as in Figure~\ref{fig:10923_zoom}. The contours in panels~B and C indicate a continuum intensity of 50 \% of the quiet Sun level. In order to understand the shape of the histogram, we select specific regions in the histogram and show examples of the corresponding features in the bottom two panels. Yellow triangles: outer parts of the umbra, red circles: spines, blue squares: transition from spines to filaments, green diamonds: dark central part of filaments, orange triangles: bright heads of filaments, and purple triangles: filaments tails exhibiting opposite polarity field. We reversed the sign of $B_{\rm z}$ to be positive in the umbra}
\label{fig:hist_examp}
\end{figure}

We study the connection between $I_{\rm C}$ and $B_{\rm z}$ and their dependence on spot size in more detail by computing 2D histograms of these two parameters. We evaluate $B_{\rm z}$ at $\log \tau = -0.9$ and we swap the sign of the magnetic field vector in cases where it points inwards in the umbra (i. e., we ensure that $B_{\rm z}$ is positive in the umbra and is negative where the flux returns below the surface). The top panel of Figure~\ref{fig:hist_examp} shows an example of such a histogram for the sunspot AR~10923. As indicated in the maps of $I_{\rm C}$ and $B_{\rm z}$ (panels~B and C in Figure~\ref{fig:hist_examp}), the various constituents of the sunspot (umbra, penumbral filaments, spines) appear at different locations in the histogram. The umbra is dark ($I_{\rm C}=0.1\--0.2$) and exhibits high field strengths (more than 2000~G). There is a smooth transition from the umbra to the spines, which have a weaker vertical magnetic field (1000 -- 2000~G) and a higher continuum intensity (up to $0.9$ of the quiet Sun). The penumbral filaments cover only a small part of the parameter space. They have a weak vertical magnetic field of a few 100~G and stretch over a narrow range in $I_{\rm C}$, from $I_{\rm C} \sim 0.5$ to $I_{\rm C} \sim 1$. The heads of the filaments are bright, but their centers are darker than the spines. The tails of the filaments are bright, too, and the polarity of $B_{\rm z}$ there is opposite to the one of the umbra. This description of the penumbral filaments is in agreement with the standard filament of \citet{2013A&A...557A..25T}. It also seems as if there is a smooth transition from spines to filaments in the histograms. It is unclear, whether this is a real signal or whether it is caused by the limited spatial resolution of the Hinode observations. The dominant part of the penumbral filaments exhibits only a very weak vertical magnetic field, though.

\begin{figure}
\centering
\resizebox{\hsize}{!}{\includegraphics{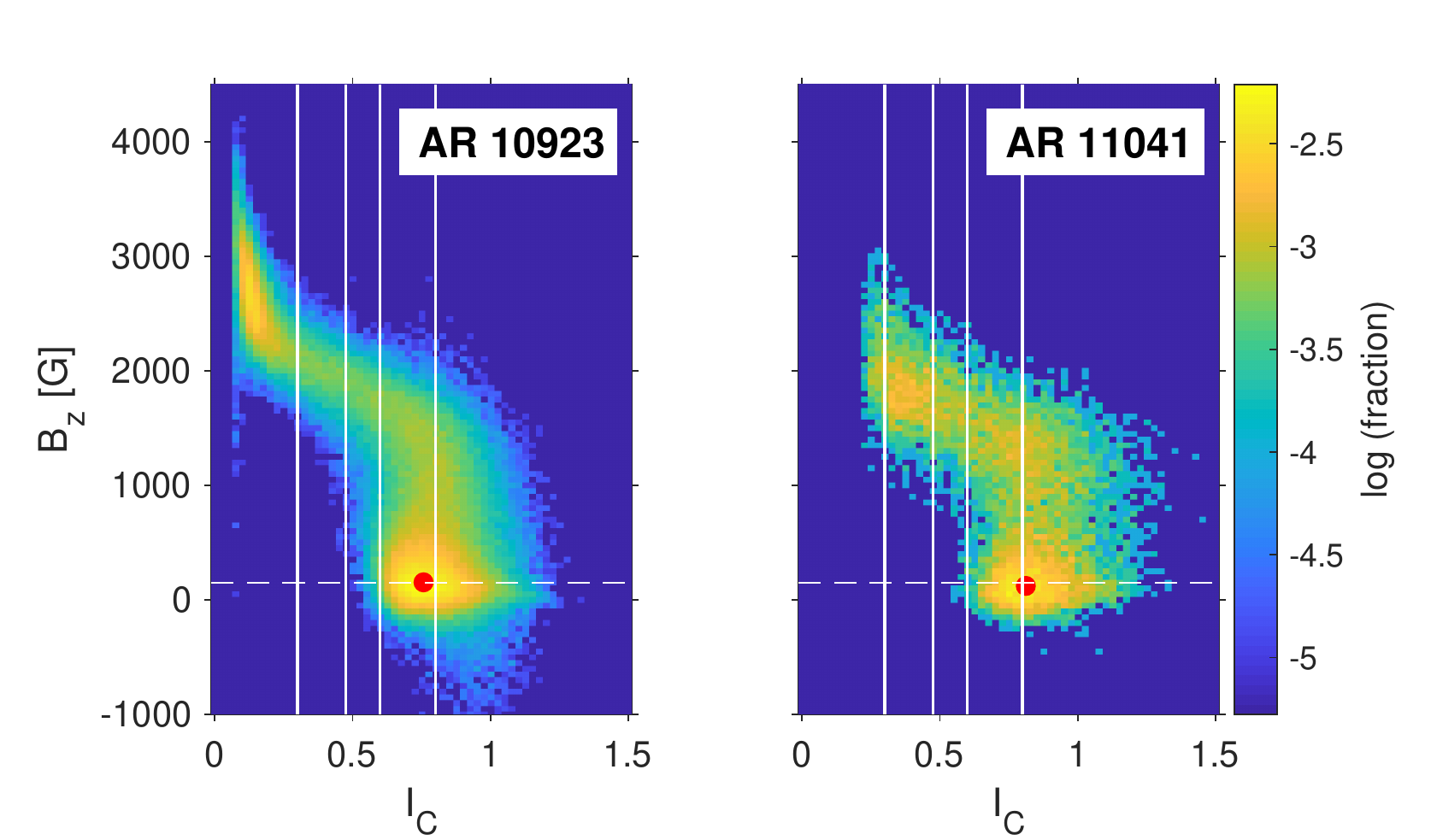}}
\caption{2D histograms of the vertical magnetic field (at $\log \tau = -0.9$) and of the continuum intensity in two sunspots, the large spot AR~10923 observed on 14 November 2006 ({\it left}) and the small spot AR~11041 observed on 26 January 2010 ({\it right}). The histograms are plotted on a logarithmic scale. The dashed horizontal line at 150~G (i. e., the penumbral filaments) indicates the position of a horizontal cut across the histograms, which are shown in the left panel of Figure~\ref{fig:hist_size2}. The vertical lines (at $I_{\rm C} = 0.30$, $0.475$, $0.60$, and $0.80$) mark the position of vertical cuts (shown in Figure~\ref{fig:hist_cuts}). The red circles show the position of the maximum of the part of the histogram that is affected by the filaments. See text for more details.}
\label{fig:hist_size1}
\end{figure}

\begin{figure}
\centering
\resizebox{\hsize}{!}{\includegraphics{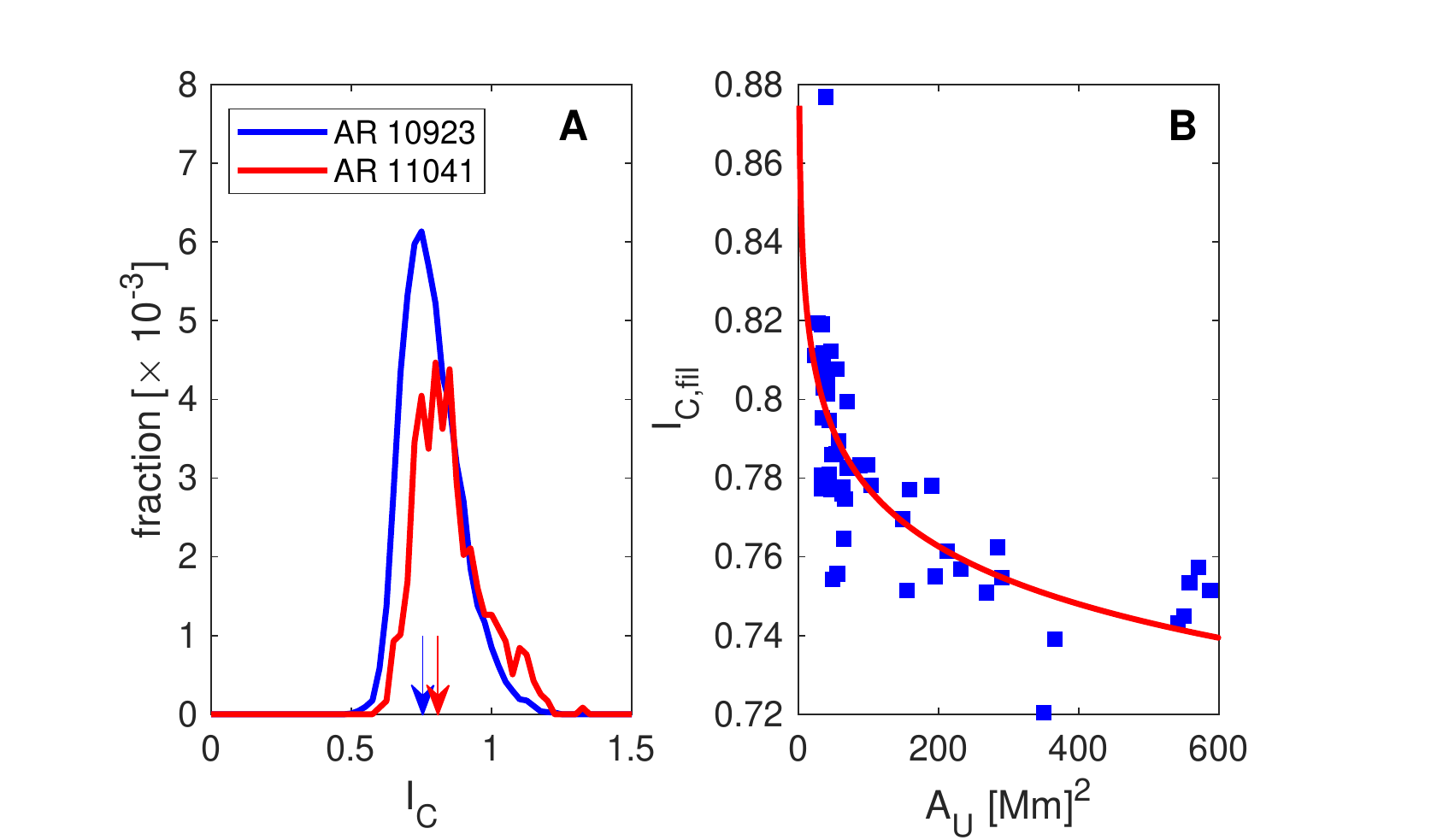}}
\caption{{\it Panel~A:} Histograms of the continuum intensity for a fixed $B_{\rm z} = 150$~G for AR~10923 (blue) and AR~11041 (red). These are horizontal cuts across the two histograms shown in Figure~\ref{fig:hist_size1} along the dashed horizontal lines. The arrows show the intensity $I_{\rm C,fil}$ corresponding to the maximum of the region of the histograms that affected by the filaments (as indicated by the red circles in Figure~\ref{fig:hist_size1}. {\it Panel~B:} $I_{\rm C,fil}$ for all the spots in our sample as a function of the total size of the umbra. The red line shows a fit to the data ($I_{\rm C,fil} = 0.91-0.02 \log{A_{\rm U}}$).}
\label{fig:hist_size2}
\end{figure}

We can now use such 2D histograms of $I_{\rm C}$ and $B_{\rm z}$ to compare these parameters between spots of different sizes. In Figure~\ref{fig:hist_size1}, we show both, the histogram of the large spot AR~10923 (umbral area 571~Mm$^2$, the same histogram as in Figure~\ref{fig:hist_examp}) and the one of a small sunspot (AR~11041, umbral area 35~Mm$^2$). The basic shape of the histogram is the same for both spots. The main difference is that the umbra of the smaller spot is brighter and has a weaker magnetic field, as expected. In addition, it also looks like the penumbral filaments are brighter in the small spot, too. Panel~A of Figure~\ref{fig:hist_size2} shows a horizontal cut across the region in the histograms that is covered by penumbral filaments (at 150~G, as indicated by the horizontal dashed lines in Figure~\ref{fig:hist_size1}). Both histograms look very similar, but the one of AR~11041 is shifted towards slighty higher continuum intensities compared to the one of AR~10923. This suggests that the brightness of the penumbral filaments in small spots is higher than of the ones in large spots. 

We analyze this in a more quantitative way by determining the location of the penumbral filaments in the $I_{\rm C}\--B_{\rm z}$ histograms of the individual spots. We derive the locations by fitting a 2D fourth-order polynomial to the region in the 2D histograms that is affected by the penumbral filaments. Examples for the position of these maxima are given by the red circles in Figure~\ref{fig:hist_size1}. Panel~B of Figure~\ref{fig:hist_size2} shows that the continuum intensity of the penumbral filaments $I_{\rm C,fil}$ estimated in this manner decreases with increasing spot size, from about $I_{\rm C} = 0.88$ for the smallest spots to $I_{\rm C}= 0.75$ for the largest ones. \citet{1992sers.conf..130B} estimated the brightness (defined as the total intensity between 500~nm and 600~nm) of entire sunspots to decrease like $-0.0244 \log{A_{\rm S}}$, where $A_{\rm S}$ is the area of the spot. We observe a similar behavior for the penumbral filaments, with the brightness depending on the area of the umbra as $I_{\rm C,fil} = 0.875-0.021 \log{A_{\rm U}}$ (see the red line in panel~B of Figure~\ref{fig:hist_size2}). The fit slightly underestimates the intensity for both, the smallest and the largest spots in our sample, though. The strength of the vertical magnetic field of the penumbral filaments does not systematically vary with spot area (mean $B_{\rm z} \sim 130$~G).

\subsection{Averaging the vertical magnetic field at a fixed intensity} \label{sect:avg}
We showed in the previous subsection that the continuum intensity and the strength of the vertical magnetic field of the constituents of sunspots vary between spots of different sizes. Now we can study how these variations with spot size affect averages of $B_{\rm z}$ at fixed values of the continuum intensity.

\begin{figure}
\centering
\resizebox{\hsize}{!}{\includegraphics{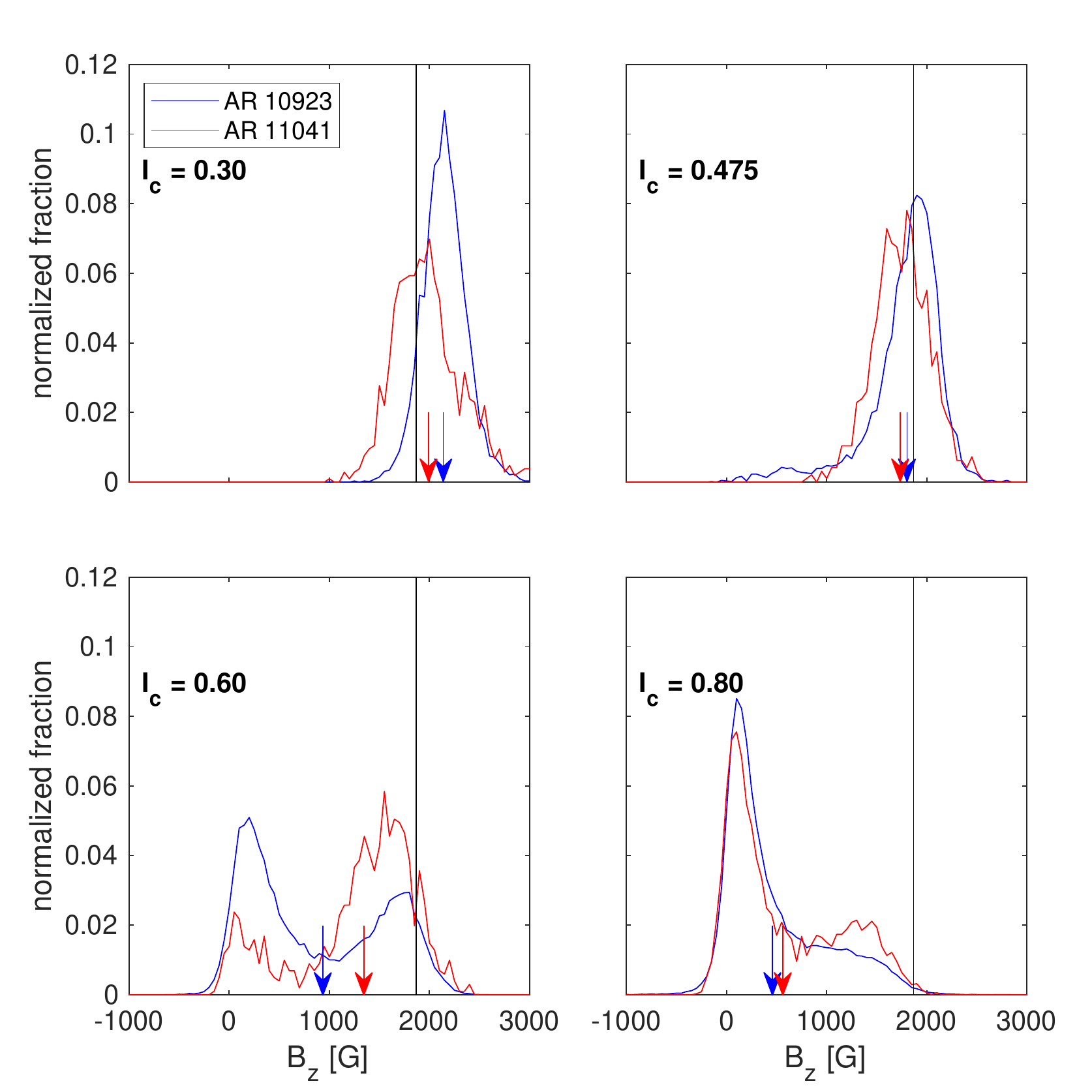}}
\caption{Vertical cuts across the 2D histograms shown in Figure~\ref{fig:hist_size1} at different values of the continuum intensity (indicated by the vertical white lines in Figure~\ref{fig:hist_size1}). {\it From top left to bottom right:} $I_{\rm C}=0.30$, $0.475$, $0.60$, and $0.80$. In all panels, the blue curve corresponds to AR~10923 and the red one to AR~11041. We normalized each histogram to unity. The arrows indicate the mean value of the vertical magnetic field for the different distributions. The black vertical line shows 1867~G for comparison.}
\label{fig:hist_cuts}
\end{figure}

Again, we start by looking at the histograms of AR~10923 and AR~11041 in Figure~\ref{fig:hist_size1}. The vertical lines in these histograms indicate selected values of the continuum intensities ($I_{\rm C} = 0.30$, $0.475$, $0.60$, and $0.80$). In Figure~\ref{fig:hist_cuts}, we show the distribution of $B_{\rm z}$ in the two histograms at these values of $I_{\rm C}$. Low intensities (e. g., $I_{\rm C} = 0.30$, top left panel in Figure~\ref{fig:hist_cuts}) occur only in the umbra. Hence, the larger spot (AR~10923) exhibits a stronger $B_{\rm z}$ at this intensity. A continuum intensity of $I_{\rm C} = 0.475$ corresponds roughly to the UP boundary. However, as discussed in Section~\ref{sect:boundary}, a threshold of $I_{\rm C}\approx 0.5$ does not outline the UP boundary consistently for both, small and large spots. Some parts of the penumbral filaments of large spots, such as AR~10923, exhibit lower continuum intensities than this threshold (see also Figure~\ref{fig:hist_size1}). These dark parts of the penumbral filaments also affect the histogram of $B_{\rm z}$ of AR~10923 at $I_{\rm C} = 0.475$, where they appear as an extended tail at low field strengths. Even though the dominant influence on the histograms still arises from the umbra and from the spines, the weak $B_{\rm z}$ of the penumbral filaments causes the mean $B_{\rm z}$ of the two spots at $I_{\rm C} = 0.475$ to be almost the same. At higher intensities ($I_{\rm C} = 0.60$ or $I_{\rm C} = 0.80$), penumbral filaments also start appearing in the histograms of AR~11041, but not as strongly as in case of AR~10923. The histograms of $B_{\rm z}$ of the large spot are dominated by penumbral filaments, which makes the mean $B_{\rm z}$ of AR~10923 lower than the one of AR~11041 at these intensities.

\begin{figure}
\centering
\resizebox{\hsize}{!}{\includegraphics{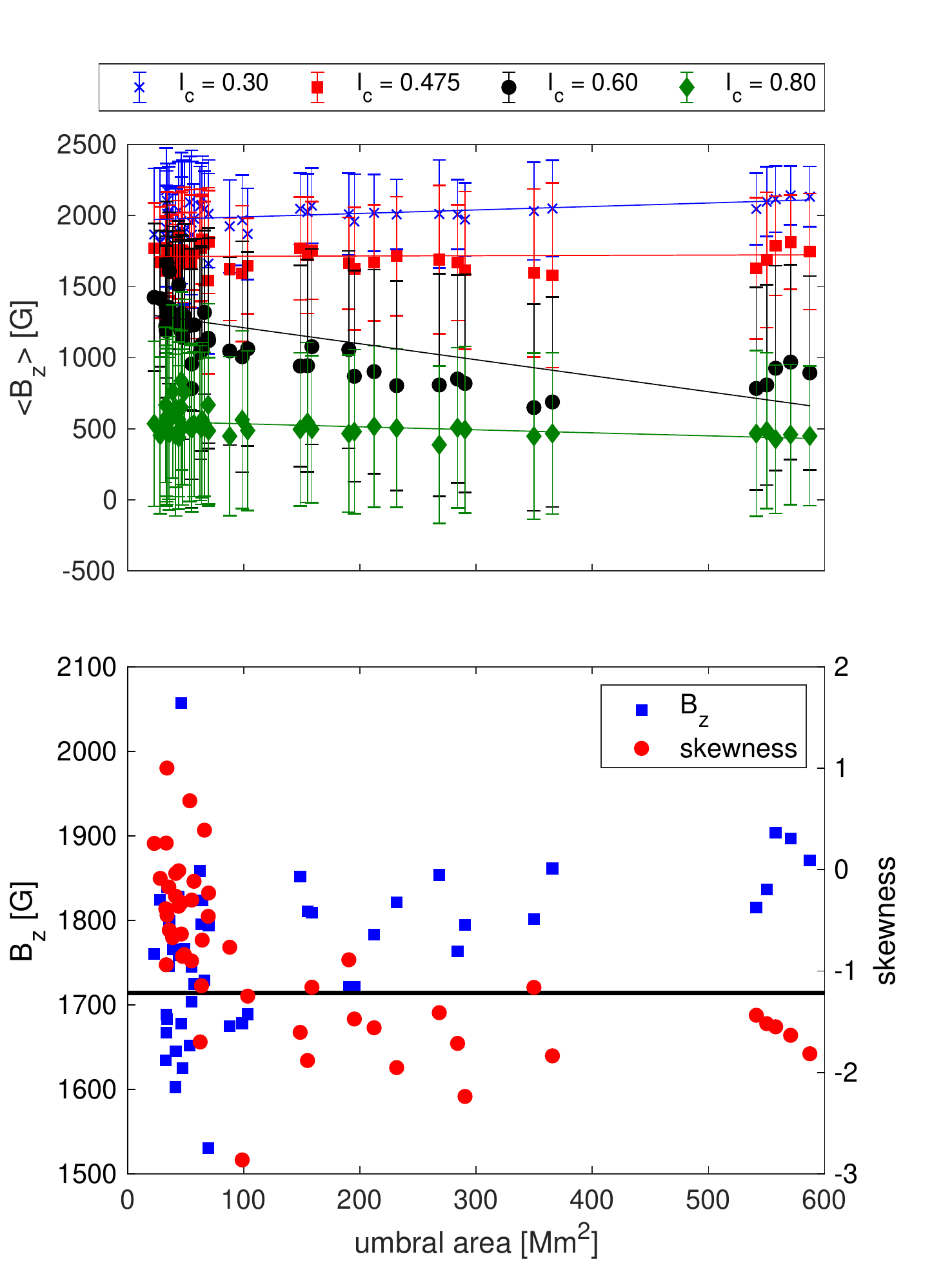}}
\caption{Dependence of the vertical magnetic field on intensity and umbral area. {\it Top panel:} vertical magnetic field averaged at fixed values of the continuum intensity plotted as a function of the area of the umbra for the individual spots. We consider four different values of the continuum intensity, $0.30$ (blue), $0.475$ (red), $0.60$ (black), and $0.80$ (green). The error bars correspond to the standard deviation of the magnetic field within the individual spots. The solid lines are linear fits for the different values of the continuum intensity. {\it Bottom panel:} estimate of the strength of $B_{\rm z}$ of the spines at $I_{\rm C} = 0.475$ (blue squares, see text for more details) and the skewness of the histograms of $B_{\rm z}$ (red circles) at this intensity as a function of the umbral area. The horizontal line indicates the mean $B_{\rm z}$ at $I_{\rm C} = 0.475$ averaged over all spots in our sample.}
\label{fig:fits}
\end{figure}

Thus, the dependence of the average $B_{\rm z}$ at a fixed continuum intensity on the size of the umbra is determined by two competing effects. On the one hand, the strength of $B_{\rm z}$ in the umbra and in spines increases with increasing spot size. On the other hand, the penumbral filaments are darker in larger spots, which means that the weak $B_{\rm z}$ of the filaments can already influence the histograms at lower intensities than for small spots. Which of these two effect dominates depends on the value of the continuum intensity, at which the average $B_{\rm z}$ is computed. Low intensities (e. g., $I_{\rm C} = 0.30$) occur only in the umbra, meaning that the average $B_{\rm z}$ at this intensity increases with umbral area (see top panel of Figure~\ref{fig:fits}). At higher intensities ($I_{\rm C} = 0.60$ or $I_{\rm C}= 0.80$), large spots start to exhibit a significant contribution from penumbral filaments, while small spots are still dominated by spines. This causes the average $B_{\rm z}$ at $I_{\rm C} = 0.60$ or $I_{\rm C} = 0.80$ to decrease with $A_{\rm U}$. 

At $I_{\rm C} \approx 0.475$, the averaged vertical magnetic field does not change with umbral area, as was already observed by \citet{2018A&A...611L...4J}. This is because the two competing influences on the dependence of the average of $B_{\rm z}$ at a fixed continuum intensity on $A_{\rm U}$ (spines: increase with $A_{\rm U}$, filaments: decrease with $A_{\rm U}$) cancel out each other at $I_{\rm C} \approx 0.5$. The strength of $B_{\rm z}$ of the spines at $I_{\rm C}= 0.475$ (estimated as the position of the peak caused by the spines in the histograms at $I_{\rm C}=0.475$) is higher for larger spots than for smaller ones, as expected (see the bottom panel of Figure~\ref{fig:fits}).

The penumbral filaments influence the shape of the histograms. Histograms that are only affected by spines and umbra are roughly symmetric. When penumbral filaments are present at $I_{\rm C}=0.475$, they distort the histogram of $B_{\rm z}$, adding an extended tail towards low field strengths, as described above. This can be quantified by computing the skewness of the histograms. As shown in the bottom panel of Figure~\ref{fig:fits}, the skewness is about zero for small spots, indicating a roughly symmetric distribution. When going towards larger spot sizes, the skewness becomes more and more negative, which means that the histograms of $B_{\rm z}$ exhibit more and more extended tails towards lower field strengths. Hence, the influence of penumbral filaments on the average $B_{\rm z}$ at $I_{\rm C} = 0.5$ increases with increasing spot size, canceling the increasing $B_{\rm z}$ of the spines.

We infer an average $B_{\rm z}$ at $I_{\rm C} = 0.475$ of $B_{\rm thr} = (1714\pm 54)$~G, which is lower than the 1867~G reported by \citet{2018A&A...611L...4J}. The differences probably arise from differences in the inversion (see next section). Indeed, the observed behavior depends on optical depth. Repeating the same analysis at $\log \tau = 0$ leads to a constant $B_{\rm z}$ of 1804~G at $I_{\rm C} = 0.45$, in much better agreement with the results of \citet{2018A&A...611L...4J}. Therefore, we can reproduce the result of \citet{2018A&A...611L...4J}, that the average $B_{\rm z}$ at $I_{\rm C} \approx 0.5$ does not depend on the size of the umbra. However, as explained above, this behavior is caused by the dependence of the brightness of penumbral filaments on $A_{\rm U}$, rather than by the onset of convection at this $B_{\rm z}$. In addition, the plots shown in Figure~\ref{fig:fits} do not reveal any differences between stable sunspots and decaying or forming ones.

We note that the error bars in the top panel of Figure~\ref{fig:fits} are clearly overestimated. The scatter between the mean $B_{\rm z}$ of the individual spots should be comparable to the size of the error bars, which is not the case in Figure~\ref{fig:fits}. This is another indication that the distribution of the vertical magnetic field at a fixed value of the continuum intensity is not well represented by its average.

\subsection{Evaluating the vertical magnetic field at different optical depths}\label{sect:tau}
\begin{table}
\caption{Derived values of $B_{\rm thr} (\log \tau)$ at the different nodes used in the inversion and the magnitude of the corresponding continuum intensity.}
\label{tab:Bconst}
\centering
\begin{tabular}{r l l}
\hline\hline
$\log \tau$ & required $I_{\rm C}$ & $B_{\rm thr}$ [G] \\
\hline
0 & $0.45$ & $1804 \pm 70$\\
$-0.9$ & $0.4750$ & $1714 \pm 54$\\
$-2.5$ & $0.525$ & $1467 \pm 50$\\
\hline
\end{tabular}
\end{table}

\begin{figure*}
\centering
\includegraphics[width=17cm]{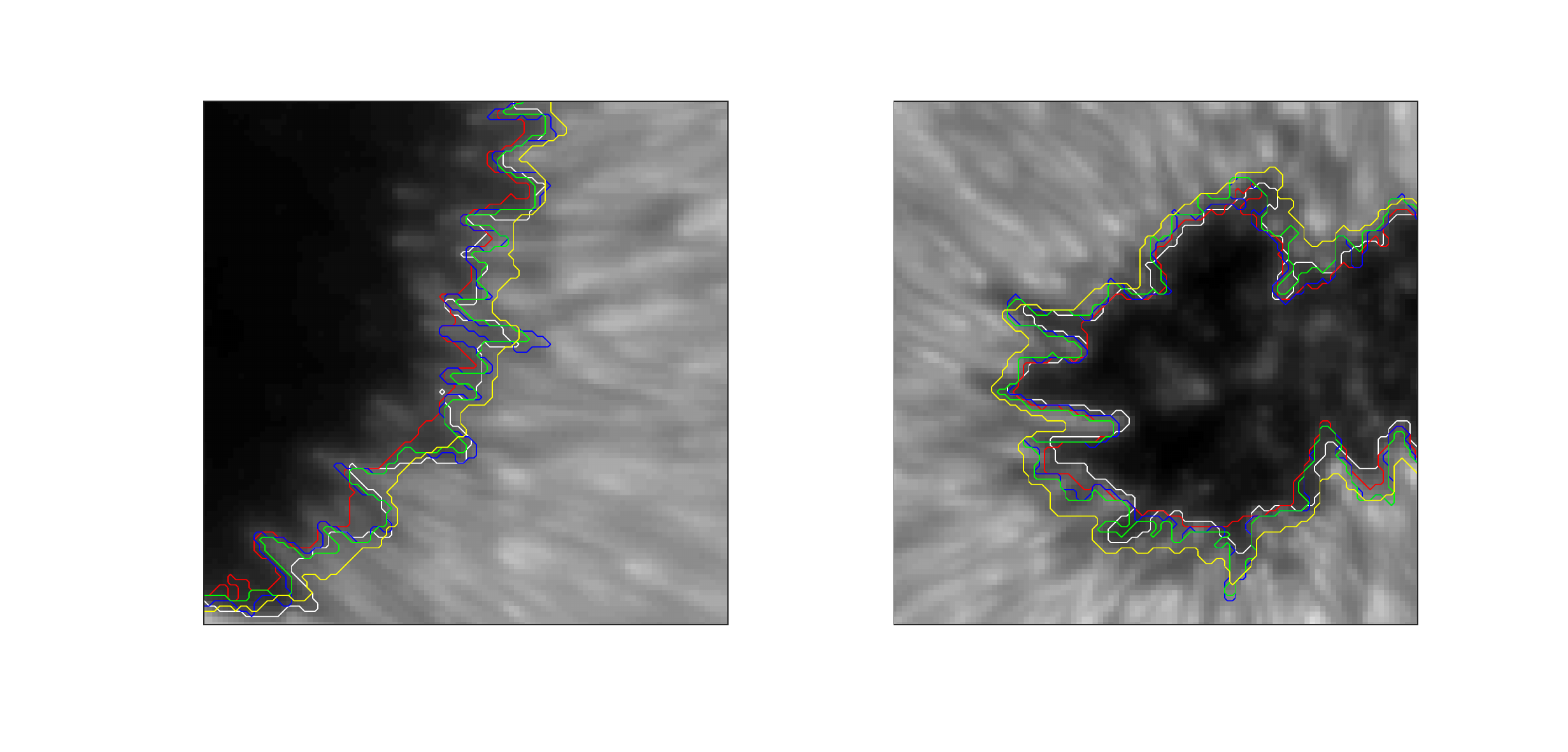}
\caption{Maps of the continuum intensity for AR~10923 observed on 14 November 2006 ({\it left}) and for AR~10969 observed on 27 August 2007 ({\it right}). The white contours indicate a continuum intensity of 50 \% of the quiet Sun level. The other contours correspond to $B_{\rm crit} (\log \tau)$ inferred for the different nodes in the inversion (see Table~\ref{tab:Bconst}). Blue: $\log \tau = 0$, green: $\log \tau = -0.9$, and yellow: $\log \tau = -2.5$. The red contour represents an inversion, where the magnetic field is independent of height, and shows a value of 1867~G. The field-of-view of the two panels is indicated by the white boxes in Figure~\ref{fig:spots}. The size of the field-of-view is the same in both panels.}
\label{fig:AR_tau}
\end{figure*}

At each node of the inversion, there is a fixed value of the continuum intensity such that the average of $B_{\rm z}$ at this intensity does not depend on the size of the sunspot. However, the values of both the derived $B_{\rm thr}$ and the continuum intensity, where it is observed, change with optical depth (see Table~\ref{tab:Bconst}). The lower the optical depth, the lower the derived $B_{\rm thr}$ and the higher the corresponding continuum intensity. Therefore, the contours of $B_{\rm thr} (\log \tau)$ are located further away from the center of the spot when going to higher atmospheric layers (see Figure~\ref{fig:AR_tau}). There is no range in optical depth, where the magnetic field outlines the UP boundary properly, though. The inconsistency in the definition of the UP boundary using $B_{\rm z}$ between sunspots of different sizes, that was discussed in Section~\ref{sect:boundary} occurs at all nodes in optical depths. At all nodes of the inversion, the strength of $B_{\rm z}$ in some penumbral filaments of large spots is higher than the $B_{\rm z}$ in the outer parts of the umbra of small spots. 

The observed increase of $B_{\rm thr} (\log \tau)$ with optical depth and the decrease of the corresponding continuum intensity are caused by the changing appearance of the penumbral filaments with height and by the decrease of the magnitude of $B_{\rm z}$ with height in the umbra and in the spines. Since the flux tube underlying the sunspot expands with height, penumbral filaments near the UP boundary appear at low optical depths to be shorter and do not extend as far into the umbra as in the deeper layers of the atmosphere. In case of large spots, the strength of $B_{\rm thr}$ is lower than the strength of $B_{\rm z}$ of the innermost parts of the penumbral filaments. Therefore, the shorter length of the penumbral filaments in the higher layers of the atmosphere requires $B_{\rm z}$ to be evaluated at a larger distance from the umbra at low optical depths, which corresponds to a higher $I_{\rm C}$. The continuum intensity increases with distance from the umbra. Since the strength of the $B_{\rm z}$ of the spines decreases with increasing distance from the umbra, the inferred $B_{\rm thr} (\log \tau)$ increases with optical depth. Some of the decrease of $B_{\rm thr} (\log \tau)$ with height is also caused by the decrease of the $B_{\rm z}$ of the spines with height in the atmosphere (we note that the $B_{\rm z}$ of penumbral filaments decreases with increasing optical depth). However, the dominant reason for the observed increase of $B_{\rm thr} (\log \tau)$ with optical depth is the changing appearance of the penumbral filaments at the different atmospheric layers.

\citet{2018A&A...611L...4J} derived a $B_{\rm thr} (\log \tau)$ of 1867~G at $I_{\rm C} = 0.5$ using Hinode observations that were inverted assuming that the magnetic field is independent of height. Consequently, their value of $B_{\rm thr} (\log \tau)$ cannot be directly compared to our results. For a better comparison, we inverted two sunspots in a similar way as done by \citet{2018A&A...611L...4J}. We performed inversions with the spatially coupled version of SPINOR and kept all parameters independent of height, except for the temperature, for which we used three nodes in optical depth placed at $\log{\tau} = -2.5,-0.9,0$. The red contours in Figure~\ref{fig:AR_tau} show the UP boundary defined by applying $B_{\rm thr} = 1867$~G to the resulting maps of $B_{\rm z}$. Again, the UP boundary cannot be defined consistently for the two spots.

\begin{figure}
\centering
\resizebox{\hsize}{!}{\includegraphics{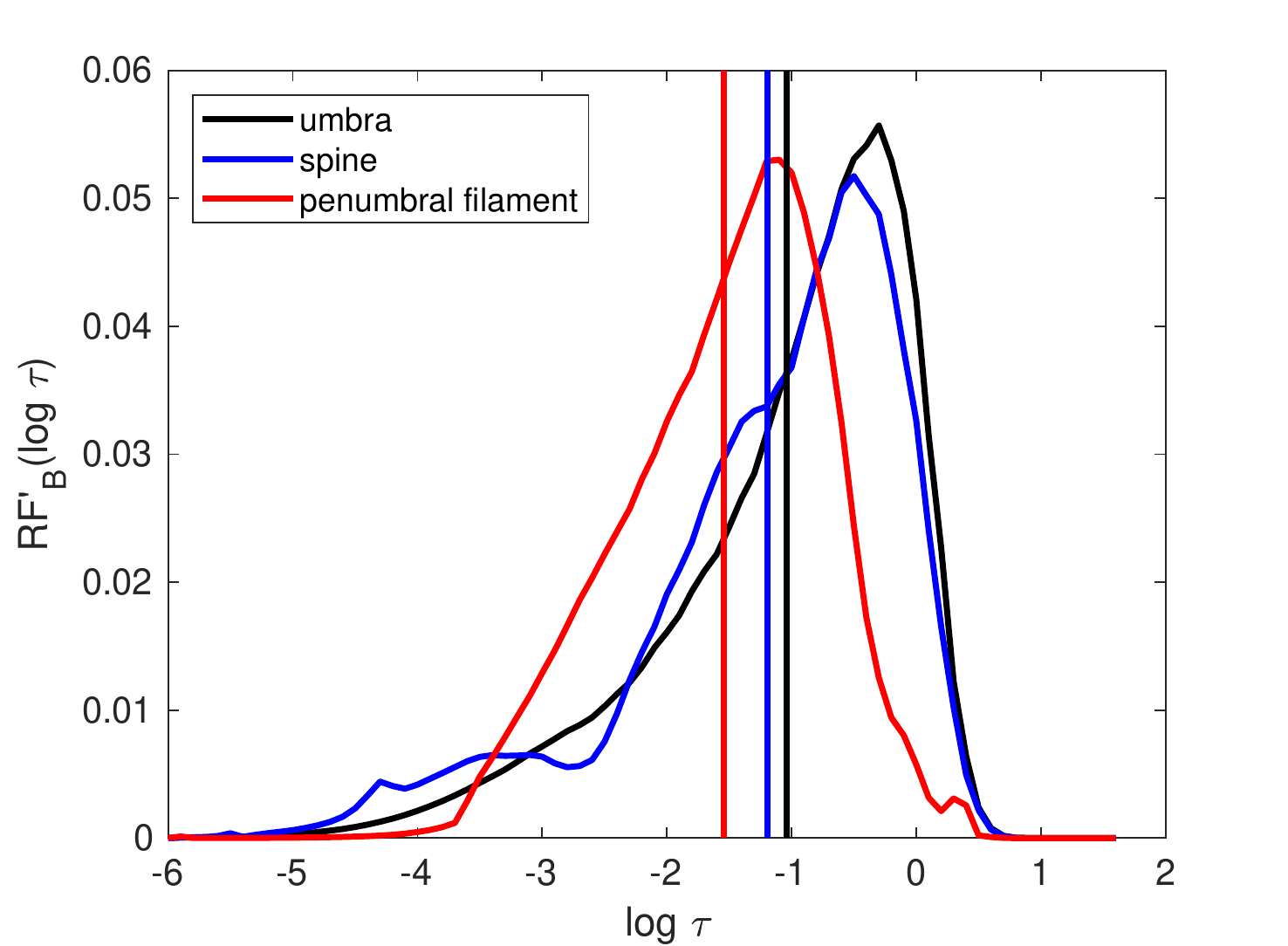}}
\caption{Representative examples of response functions for the strength of the magnetic field computed from a MHD simulation of a sunspot provided by \citet{2012ApJ...750...62R}. Black: umbra, blue: spine, red: penumbral filament. The vertical lines indicate the center-of-gravity of the corresponding response functions. See text for more details.}
\label{fig:RFs}
\end{figure}

When assuming the magnetic field to be independent of height in the inversion, the derived $B_{\rm z}$ does not correspond to a fixed optical depth. Instead, the inverted magnetic field is affected by a broad range in optical depth, depending on the formation height of the spectral lines that were observed. The range in optical depth that can influence the inverted magnetic field is given by the response function of the observed spectral lines. We computed response functions for the magnetic field strength for the various constituents of a sunspot. The response functions are based on representative atmospheres for the umbra, penumbral filaments, and spines, that were extracted from an MHD simulation of a sunspot computed by \citet{2012ApJ...750...62R} and were computed using SPINOR. We define the response function for the magnetic field strength $\textmd{RF}_{\rm B}(\lambda,\log \tau)$ as $\delta I(\lambda) = \int_{-\infty}^{\infty} \textmd{RF}_{\rm B} (\lambda,\log \tau) \delta B(\log \tau) \textmd{d} \log \tau$.  We then integrate the absolute value of the response functions over wavelength and normalize them by their integral over $\log \tau$ for better visibility:
\begin{align}
\textmd{RF}_{\rm B}'(\log \tau) = \frac{\int |\textmd{RF}_{\rm B}(\lambda,\log \tau)| \textmd{d} \lambda}{\int \int |\textmd{RF}_{\rm B}(\lambda,\log \tau)| \textmd{d} \lambda \textmd{d} \log \tau}.
\end{align}
In Figure~\ref{fig:RFs}, we show the resulting $\textmd{RF}_{\rm B}'(\log \tau)$ for the umbra, spines, and penumbral filaments, respectively. The response functions of the umbra and the spine look very similar, with a higher sensitivity to the magnetic field in the deeper layers of the atmosphere. The center-of-gravity of $\textmd{RF}_{\rm B}'$ lies at $\log \tau = -1.0$ (umbra) or $\log \tau = -1.2$ (spine), respectively. In penumbral filaments, the inversion is more sensitive to the magnetic field in higher atmospheric layers (center-of-gravity of $\textmd{RF}_{\rm B}'$ at $\log \tau = -1.5$).

Due to these differences in the formation height of the Fe~I lines across the sunspot, the $B_{\rm thr} (\log \tau)$ of 1867~G obtained by \citet{2018A&A...611L...4J} cannot be directly compared to our results of $B_{\rm thr} (\log \tau)$ at a fixed optical depth. Even though the center-of-gravity of the response functions is not too far away from the center node of our inversion ($\log \tau = -0.9$), the derived values of $B_{\rm thr}$ differ significantly.


\section{The decaying spot AR~10953}
\begin{figure}
\centering
\resizebox{\hsize}{!}{\includegraphics{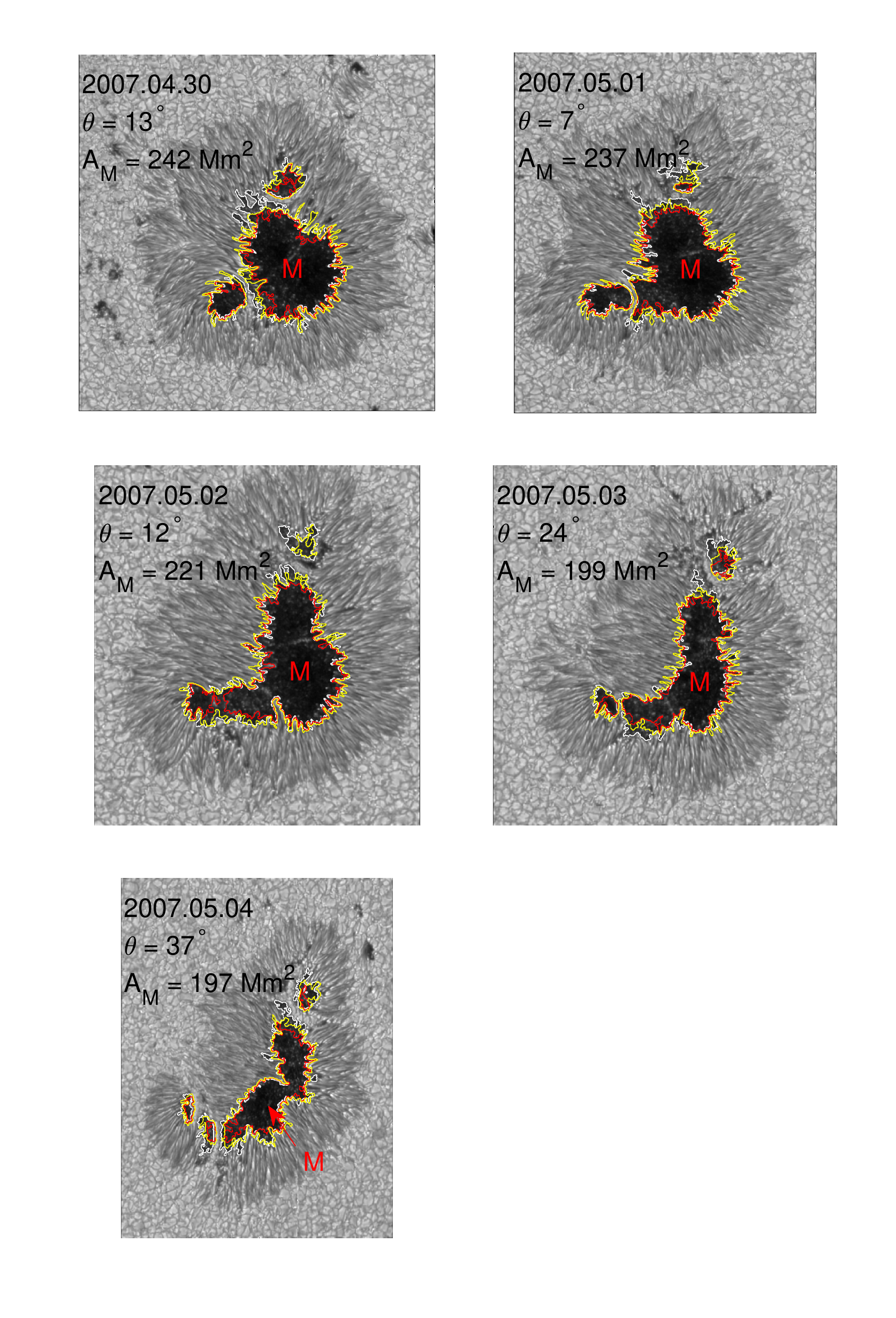}}
\caption{Evolution of AR~10953 over five consecutive days. This sunspot was observed by Hinode every day from 30 April 2007 to 4 May 2007. The white contours indicate a continuum intensity of 50 \% of the quiet Sun level, the yellow contour shows a $B_{\rm z} = 1714$~G evaluated at $\log \tau = -0.9$ (the threshold for the \jurcak{} criterion for our data, see Section~\ref{sect:avg}. For comparison, we also indicate in red a value of $B_{\rm z} = 1867$~G (the original value for the threshold for the \jurcak{} criterion). We are interested in the UP boundary of the main umbral core of this spot, which is highlighted by the label M in all panels.}
\label{fig:10953}
\end{figure}

\citet{2018A&A...611L...4J} interpreted the parts of the umbra where $B_{\rm z}$ is lower than given by the \jurcak{} criterion as to be about to be converted into penumbra. Correspondingly, parts of the penumbra with a stronger $B_{\rm z}$ than given by the \jurcak{} criterion should be stable against convection and should be converted into umbra.

One of the spots in our sample (AR~10953) was observed by Hinode for five consecutive days (from 30 April 2007 to 4 May 2007, see Figure~\ref{fig:10953}). During the entire time, the UP boundary defined using $B_{\rm z} = 1714$~G (the threshold for the \jurcak{} criterion for our data, see previous section) lies well within the penumbra (yellow contour in Figure~\ref{fig:10953}). We do not see any indications that parts of the penumbra of the spot get converted into umbra. To the contrary, the size of the main umbral core of the sunspot (indicated by the M in Figure~\ref{fig:10953}) even decreases by about 20\% within the five days. A higher threshold of $B_{\rm z} = 1867$~G \citep[the threshold given by][]{2018A&A...611L...4J} does not falsely assign the UP boundary to lie within the penumbra and is in reasonably good agreement with the $I_{\rm C} = 0.5$ contour. However, following the arguments of \citet{2018A&A...611L...4J}, the UP boundary given by the \jurcak{} criterion should be located within the umbra. The area of the umbra decreases over the observed period in time. Hence, parts of the umbra are about to be converted to penumbra and thus, should violate the \jurcak{} criterion. Therefore, an even higher threshold for $B_{\rm z}$ is needed to describe the evolution of the UP boundary in AR~10953. Such a high threshold is not supported by our results from Section~\ref{sect:avg}.

The failure of the \jurcak{} criterion in predicting the decay of the umbra of this sunspot is not caused by the setup of the inversion. We tested this by inverting the Hinode observations of AR~10953 on 1 May 2007 with the spatially coupled version of SPINOR under the assumption that the magnetic field vector does not change with optical depth. This setup of the inversion corresponds closely to the one used by \citet{2018A&A...611L...4J}. Even in this case, a contour of $B_{\rm z}$ assigns some parts of penumbral filaments to be part of the umbra, the \jurcak{} criterion does not predict the decrease of the area of the umbra.

The fact that the UP boundary does neither adjust to the contour given by $B_{\rm z} = 1714$~G or to the one given by $B_{\rm z} = 1867$~G therefore challenges the interpretation of the \jurcak{} criterion as to be related to the onset of magnetoconvection.

\section{Discussion}
We have shown that there is no unique value of $B_{\rm z}$ which can outline the UP boundary consistently in all sunspots that we have analyzed. The smaller the spot, the lower the strength of the vertical magnetic field at the UP boundary. \citet{2018A&A...611L...4J} interpreted a failure of the \jurcak{} criterion to be related to the conversion of umbra to penumbra. This interpretation was supported by the observations of a decaying sunspot by \citet{2018A&A...620A.191B}, where the $B_{\rm z}$ is lower than $B_{\rm thr}$ in parts of the umbra. Indeed, many of the sunspots in our sample, where the \jurcak{} criterion is violated, are decaying spots. However, we see some indications that this theory is not correct. We observe that the strength of $B_{\rm z}$ at the UP boundary depends predominantly on the size of the spots. Failures of the \jurcak{} criterion in outlining the UP boundary occur mostly for small spots, independent of their evolutionary state. There are no obvious differences in the properties of the UP boundary between stable sunspots and decaying or forming ones of similar size. Our case study of AR~10953 also suggests that a failure of the \jurcak{} criterion is not directly related to sunspot decay. Even though the area of the umbra of this large sunspot decreases significantly over the course of a few days, the \jurcak{} criterion predicts a UP boundary that lies within the penumbra during the entire time period.

The magnetic field at the UP boundary has been studied using several instruments, such as Hinode/SOT \citep[][and this work]{2018A&A...611L...4J}, HMI \citep{2018A&A...620A.104S}, and ground-based observations from GREGOR \citep{2020arXiv200409956L}. A canonical value of $B_{\rm z}$ was observed in all of these studies, although the exact value of the inferred $B_{\rm thr}$ differs. This is probably due to observations of different spectral lines and differences in the inversion (see also Section~\ref{sect:tau}). Our study is based only on one instrument and one spectral line. However, throughout the entire range in optical depth, we can explain the critical value of $B_{\rm thr}$ best associated with the umbra-penumbra boundary of a given spot by the properties of penumbral filaments and spines and their dependence on spot size. Although the exact value of $B_{\rm thr}$ can depend on the employed spectral line, the physical explanation for the dependence of $B_{\rm thr}$ on sunspot area should be independent of it. Therefore, we expect our interpretation of the \jurcak{} criterion to apply to data provided by other instruments, as well.

A dependence of the properties of the magnetic field at the UP boundary on the size of the spot could also be an alternative explanation for the observations of \citet{2018A&A...620A.191B}. Instead of being directly caused by the decay of the spot, the decrease of $B_{\rm z}$ at the UP boundary, which they observe, might be due to the shrinking area of the sunspot, but with the $B_{\rm z}$ at the UP boundary being consistent to the one of stable sunspots of similar sizes. More generally, it is possible that the observed reconfiguration of the magnetic field in the penumbra during sunspot decay \citep{2014ApJ...796...77W,2018A&A...614A...2V,2018A&A...620A.191B} is not directly related to spot decay, either. The magnetic field of a decaying penumbra might exhibit the same properties as the one of a stable sunspot of a similar size. We note that the time scale of sunspot decay (usually a few days) is much longer than the dynamical time scale of sunspots (about one hour, the time it takes a magnetoacoustic wave to travel across a sunspot). This might suggest that the large-scale structure of sunspots is more or less in equilibrium during the decay phase. However, addressing this question requires a more detailed comparison of the structure of the magnetic field in the penumbra between stable and decaying sunspots of similar sizes, which is beyond the scope of this study.

In addition, the onset of magnetoconvection does not seem to occur at a fixed threshold of $B_{\rm z}$. The observed constant value of the average $B_z$ at $I_{\rm C} \approx 0.5$ is caused by the dependence of the brightness of penumbral filaments on the spot size. A definition of the UP boundary as where $I_{\rm C} = 0.5$ is not consistently valid for spots of different sizes, because it assigns parts of the penumbral filaments of large spots to be part of the umbra. Hence, it is not meaningful to compare the average $B_{\rm z}$ at $I_{\rm C} \approx 0.5$ between different spots. Moreover, this average $B_{\rm z}$ does not correspond to the magnetic field of a particular component of the sunspot. For small spots, the average $B_{\rm z}$ is similar to the field strength of the outer parts of the umbra and of the spines. In large spots, it lies somewhere between the field strength of the spines and of the penumbral filaments. If this value of $B_{\rm z}$ was related to the onset of magnetoconvection, we would expect a significant fraction of the $B_{\rm z}$ at the UP boundary to be around this value. Furthermore, the position of the UP boundary defined using $B_{\rm thr} (\log \tau)$ depends on optical depth. For the top node in our inversion, the contour of the $B_{\rm thr}$ derived from the magnetic field at that optical depth lies well within the penumbra. This dependence of the UP boundary defined using $B_{\rm thr} (\log \tau)$ on optical depth means that the vertical magnetic field cannot even outline the UP boundary consistently within individual sunspots. Moreover, ascribing the onset of magnetoconvection in the penumbra to the properties of the magnetic field in the photosphere is at odds with recent observations, which indicate that the magnetic field in the chromosphere plays as major role in the formation of the penumbra \citep{2012ApJ...747L..18S,2013ApJ...769L..18L,2013ApJ...771L...3R,2014ApJ...784...10R,2016ApJ...825...75M}.

It therefore remains an open question what causes the transition from umbra to penumbra and how the UP boundary can be identified. The physical properties of the umbra and spines are very similar and vary smoothly with distance from the center of the spot. This makes the distinction between these two constituents of sunspots and therefore the definition of the UP boundary somewhat arbitrary. Only the penumbral filaments can be clearly distinguished from the spines in which they are embedded. Since the penumbral filaments are the main convective features in the penumbra, their inner boundary may serve as the physically most relevant boundary between umbra and penumbra.

Our results have important implications for understanding the nature of magnetoconvection in sunspots. We have shown that there is no fixed threshold for $B_{\rm z}$ at which the umbra gets transformed into penumbra. This does not agree with the theoretical predictions made by \citet{2019ApJ...873L..10M}, who claimed that convection would be inhibited in sunspots where the vertical magnetic field exceeds a fixed threshold. However, this study was based on the stability criterion of \citet{1966MNRAS.133...85G}, which is only valid for very simple configurations of the magnetic field. As stated in \citet{1966MNRAS.133...85G}, this stability criterion cannot be applied to sunspots, which have a very complex magnetic field.

\begin{acknowledgements}
We thank the unknown referee for the useful suggestions. We are grateful to Matthias Rempel for providing the 3D MHD simulations of a sunspot. This work benefited from the Hinode sunspot database at MPS, created by Gautam Narayan. This project has received funding from the European Research Council (ERC) under the European Union’s Horizon 2020 research and innovation programme (grant agreement No 695075) and has been supported by the BK21 plus program through the National Research Foundation (NRF) funded by the Ministry of Education of Korea. Hinode is a Japanese mission developed and launched by ISAS/JAXA, collaborating with NAOJ as a domestic partner, NASA and STFC (UK) as international partners. Scientific operation of the Hinode mission is conducted by the Hinode science team organized at ISAS/JAXA. This team mainly consists of scientists from institutes in the partner countries. Support for the post-launch operation is provided by JAXA and NAOJ (Japan), STFC (U.K.), NASA, ESA, and NSC (Norway).
\end{acknowledgements}

\bibliographystyle{aa} 
\bibliography{literature} 

\begin{thebibliography}{57}
\expandafter\ifx\csname natexlab\endcsname\relax\def\natexlab#1{#1}\fi

\bibitem[{{Beck} \& {Chapman}(1993)}]{1993SoPh..146...49B}
{Beck}, J.~G. \& {Chapman}, G.~A. 1993, \solphys, 146, 49

\bibitem[{{Bellot Rubio} {et~al.}(2008){Bellot Rubio}, {Tritschler}, \&
  {Mart{\'\i}nez Pillet}}]{2008ApJ...676..698B}
{Bellot Rubio}, L.~R., {Tritschler}, A., \& {Mart{\'\i}nez Pillet}, V. 2008,
  \apj, 676, 698

\bibitem[{{Benko} {et~al.}(2018){Benko}, {Gonz{\'a}lez Manrique}, {Balthasar},
  {G{\"o}m{\"o}ry}, {Kuckein}, \& {Jur{\v c}{\'a}k}}]{2018A&A...620A.191B}
{Benko}, M., {Gonz{\'a}lez Manrique}, S.~J., {Balthasar}, H., {et~al.} 2018,
  \aap, 620, A191

\bibitem[{{Brandt} {et~al.}(1992){Brandt}, {Schmidt}, \&
  {Steinegger}}]{1992sers.conf..130B}
{Brandt}, P.~N., {Schmidt}, W., \& {Steinegger}, M. 1992, in Solar
  Electromagnetic Radiation Study for Solar Cycle 22, ed. R.~F. {Donnelly}, 130

\bibitem[{{Brummell} {et~al.}(2008){Brummell}, {Tobias}, {Thomas}, \&
  {Weiss}}]{2008ApJ...686.1454B}
{Brummell}, N.~H., {Tobias}, S.~M., {Thomas}, J.~H., \& {Weiss}, N.~O. 2008,
  \apj, 686, 1454

\bibitem[{{Chapman} {et~al.}(1994){Chapman}, {Cookson}, \&
  {Dobias}}]{1994ApJ...432..403C}
{Chapman}, G.~A., {Cookson}, A.~M., \& {Dobias}, J.~J. 1994, \apj, 432, 403

\bibitem[{{Frutiger} {et~al.}(2000){Frutiger}, {Solanki}, {Fligge}, \&
  {Bruls}}]{2000A&A...358.1109F}
{Frutiger}, C., {Solanki}, S.~K., {Fligge}, M., \& {Bruls}, J.~H.~M.~J. 2000,
  \aap, 358, 1109

\bibitem[{{Georgoulis}(2005)}]{2005ApJ...629L..69G}
{Georgoulis}, M.~K. 2005, \apjl, 629, L69

\bibitem[{{Gough} \& {Tayler}(1966)}]{1966MNRAS.133...85G}
{Gough}, D.~O. \& {Tayler}, R.~J. 1966, \mnras, 133, 85

\bibitem[{{Ichimoto} {et~al.}(2008){Ichimoto}, {Lites}, {Elmore}, {Suematsu},
  {Tsuneta}, {Katsukawa}, {Shimizu}, {Shine}, {Tarbell}, {Title}, {Kiyohara},
  {Shinoda}, {Card}, {Lecinski}, {Streander}, {Nakagiri}, {Miyashita},
  {Noguchi}, {Hoffmann}, \& {Cruz}}]{2008SoPh..249..233I}
{Ichimoto}, K., {Lites}, B., {Elmore}, D., {et~al.} 2008, \solphys, 249, 233

\bibitem[{{Jurcak} {et~al.}(2020){Jurcak}, {Schmassmann}, {Rempel}, {Bello
  Gonzalez}, \& {Schlichenmaier}}]{2020arXiv200403940J}
{Jurcak}, J., {Schmassmann}, M., {Rempel}, M., {Bello Gonzalez}, N., \&
  {Schlichenmaier}, R. 2020, arXiv e-prints, arXiv:2004.03940

\bibitem[{{Jur{\v c}{\'a}k}(2011)}]{2011A&A...531A.118J}
{Jur{\v c}{\'a}k}, J. 2011, \aap, 531, A118

\bibitem[{{Jur{\v c}{\'a}k} {et~al.}(2015){Jur{\v c}{\'a}k}, {Bello
  Gonz{\'a}lez}, {Schlichenmaier}, \& {Rezaei}}]{2015A&A...580L...1J}
{Jur{\v c}{\'a}k}, J., {Bello Gonz{\'a}lez}, N., {Schlichenmaier}, R., \&
  {Rezaei}, R. 2015, \aap, 580, L1

\bibitem[{{Jur{\v c}{\'a}k} {et~al.}(2017){Jur{\v c}{\'a}k}, {Bello
  Gonz{\'a}lez}, {Schlichenmaier}, \& {Rezaei}}]{2017A&A...597A..60J}
{Jur{\v c}{\'a}k}, J., {Bello Gonz{\'a}lez}, N., {Schlichenmaier}, R., \&
  {Rezaei}, R. 2017, \aap, 597, A60

\bibitem[{{Jur{\v c}{\'a}k} {et~al.}(2018){Jur{\v c}{\'a}k}, {Rezaei},
  {Gonz{\'a}lez}, {Schlichenmaier}, \& {Vomlel}}]{2018A&A...611L...4J}
{Jur{\v c}{\'a}k}, J., {Rezaei}, R., {Gonz{\'a}lez}, N.~B., {Schlichenmaier},
  R., \& {Vomlel}, J. 2018, \aap, 611, L4

\bibitem[{{Kiess} {et~al.}(2014){Kiess}, {Rezaei}, \&
  {Schmidt}}]{2014A&A...565A..52K}
{Kiess}, C., {Rezaei}, R., \& {Schmidt}, W. 2014, \aap, 565, A52

\bibitem[{{Kitai} {et~al.}(2014){Kitai}, {Watanabe}, \&
  {Otsuji}}]{2014PASJ...66S..11K}
{Kitai}, R., {Watanabe}, H., \& {Otsuji}, K. 2014, \pasj, 66, S11

\bibitem[{{Kopp} \& {Rabin}(1992)}]{1992SoPh..141..253K}
{Kopp}, G. \& {Rabin}, D. 1992, \solphys, 141, 253

\bibitem[{{Kosugi} {et~al.}(2007){Kosugi}, {Matsuzaki}, {Sakao}, {Shimizu},
  {Sone}, {Tachikawa}, {Hashimoto}, {Minesugi}, {Ohnishi}, {Yamada}, {Tsuneta},
  {Hara}, {Ichimoto}, {Suematsu}, {Shimojo}, {Watanabe}, {Shimada}, {Davis},
  {Hill}, {Owens}, {Title}, {Culhane}, {Harra}, {Doschek}, \&
  {Golub}}]{2007SoPh..243....3K}
{Kosugi}, T., {Matsuzaki}, K., {Sakao}, T., {et~al.} 2007, \solphys, 243, 3

\bibitem[{{Leka} \& {Skumanich}(1998)}]{1998ApJ...507..454L}
{Leka}, K.~D. \& {Skumanich}, A. 1998, \apj, 507, 454

\bibitem[{{Lim} {et~al.}(2013){Lim}, {Yurchyshyn}, {Goode}, \&
  {Cho}}]{2013ApJ...769L..18L}
{Lim}, E.-K., {Yurchyshyn}, V., {Goode}, P., \& {Cho}, K.-S. 2013, \apjl, 769,
  L18

\bibitem[{{Lindner} {et~al.}(2020){Lindner}, {Schlichenmaier}, \& {Bello
  Gonz{\'a}lez}}]{2020arXiv200409956L}
{Lindner}, P., {Schlichenmaier}, R., \& {Bello Gonz{\'a}lez}, N. 2020, arXiv
  e-prints, arXiv:2004.09956

\bibitem[{{Lites} {et~al.}(2013){Lites}, {Akin}, {Card}, {Cruz}, {Duncan},
  {Edwards}, {Elmore}, {Hoffmann}, {Katsukawa}, {Katz}, {Kubo}, {Ichimoto},
  {Shimizu}, {Shine}, {Streander}, {Suematsu}, {Tarbell}, {Title}, \&
  {Tsuneta}}]{2013SoPh..283..579L}
{Lites}, B.~W., {Akin}, D.~L., {Card}, G., {et~al.} 2013, \solphys, 283, 579

\bibitem[{{Lites} {et~al.}(1993){Lites}, {Elmore}, {Seagraves}, \&
  {Skumanich}}]{1993ApJ...418..928L}
{Lites}, B.~W., {Elmore}, D.~F., {Seagraves}, P., \& {Skumanich}, A.~P. 1993,
  \apj, 418, 928

\bibitem[{{Livingston}(2002)}]{2002SoPh..207...41L}
{Livingston}, W. 2002, \solphys, 207, 41

\bibitem[{{L{\"o}ptien} {et~al.}(2018){L{\"o}ptien}, {Lagg}, {van Noort}, \&
  {Solanki}}]{2018A&A...619A..42L}
{L{\"o}ptien}, B., {Lagg}, A., {van Noort}, M., \& {Solanki}, S.~K. 2018, \aap,
  619, A42

\bibitem[{{Mathew} {et~al.}(2007){Mathew}, {Mart{\'{\i}}nez Pillet}, {Solanki},
  \& {Krivova}}]{2007A&A...465..291M}
{Mathew}, S.~K., {Mart{\'{\i}}nez Pillet}, V., {Solanki}, S.~K., \& {Krivova},
  N.~A. 2007, \aap, 465, 291

\bibitem[{{Mullan} \& {MacDonald}(2019)}]{2019ApJ...873L..10M}
{Mullan}, D.~J. \& {MacDonald}, J. 2019, \apjl, 873, L10

\bibitem[{{Murabito} {et~al.}(2016){Murabito}, {Romano}, {Guglielmino},
  {Zuccarello}, \& {Solanki}}]{2016ApJ...825...75M}
{Murabito}, M., {Romano}, P., {Guglielmino}, S.~L., {Zuccarello}, F., \&
  {Solanki}, S.~K. 2016, \apj, 825, 75

\bibitem[{{Rempel}(2011)}]{2011ApJ...729....5R}
{Rempel}, M. 2011, \apj, 729, 5

\bibitem[{{Rempel}(2012)}]{2012ApJ...750...62R}
{Rempel}, M. 2012, \apj, 750, 62

\bibitem[{{Rempel} \& {Cheung}(2014)}]{2014ApJ...785...90R}
{Rempel}, M. \& {Cheung}, M.~C.~M. 2014, \apj, 785, 90

\bibitem[{{Rempel} {et~al.}(2009){Rempel}, {Sch{\"u}ssler}, \&
  {Kn{\"o}lker}}]{2009ApJ...691..640R}
{Rempel}, M., {Sch{\"u}ssler}, M., \& {Kn{\"o}lker}, M. 2009, \apj, 691, 640

\bibitem[{{Rezaei} {et~al.}(2015){Rezaei}, {Beck}, {Lagg}, {Borrero},
  {Schmidt}, \& {Collados}}]{2015A&A...578A..43R}
{Rezaei}, R., {Beck}, C., {Lagg}, A., {et~al.} 2015, \aap, 578, A43

\bibitem[{{Rezaei} {et~al.}(2012){Rezaei}, {Beck}, \&
  {Schmidt}}]{2012A&A...541A..60R}
{Rezaei}, R., {Beck}, C., \& {Schmidt}, W. 2012, \aap, 541, A60

\bibitem[{{Romano} {et~al.}(2013){Romano}, {Frasca}, {Guglielmino}, {Ermolli},
  {Tritschler}, {Reardon}, \& {Zuccarello}}]{2013ApJ...771L...3R}
{Romano}, P., {Frasca}, D., {Guglielmino}, S.~L., {et~al.} 2013, \apjl, 771, L3

\bibitem[{{Romano} {et~al.}(2014){Romano}, {Guglielmino}, {Cristaldi},
  {Ermolli}, {Falco}, \& {Zuccarello}}]{2014ApJ...784...10R}
{Romano}, P., {Guglielmino}, S.~L., {Cristaldi}, A., {et~al.} 2014, \apj, 784,
  10

\bibitem[{{Rucklidge} {et~al.}(1995){Rucklidge}, {Schmidt}, \&
  {Weiss}}]{1995MNRAS.273..491R}
{Rucklidge}, A.~M., {Schmidt}, H.~U., \& {Weiss}, N.~O. 1995, \mnras, 273, 491

\bibitem[{{Ruiz Cobo} \& {del Toro Iniesta}(1992)}]{1992ApJ...398..375R}
{Ruiz Cobo}, B. \& {del Toro Iniesta}, J.~C. 1992, \apj, 398, 375

\bibitem[{{Schad}(2014)}]{2014SoPh..289.1477S}
{Schad}, T.~A. 2014, \solphys, 289, 1477

\bibitem[{{Scharmer} {et~al.}(2008){Scharmer}, {Nordlund}, \&
  {Heinemann}}]{2008ApJ...677L.149S}
{Scharmer}, G.~B., {Nordlund}, {\AA}., \& {Heinemann}, T. 2008, \apjl, 677,
  L149

\bibitem[{{Schlichenmaier} {et~al.}(2010){Schlichenmaier}, {Rezaei}, {Bello
  Gonz{\'a}lez}, \& {Waldmann}}]{2010A&A...512L...1S}
{Schlichenmaier}, R., {Rezaei}, R., {Bello Gonz{\'a}lez}, N., \& {Waldmann},
  T.~A. 2010, \aap, 512, L1

\bibitem[{{Schmassmann} {et~al.}(2018){Schmassmann}, {Schlichenmaier}, \&
  {Bello Gonz{\'a}lez}}]{2018A&A...620A.104S}
{Schmassmann}, M., {Schlichenmaier}, R., \& {Bello Gonz{\'a}lez}, N. 2018,
  \aap, 620, A104

\bibitem[{{Shimizu} {et~al.}(2012){Shimizu}, {Ichimoto}, \&
  {Suematsu}}]{2012ApJ...747L..18S}
{Shimizu}, T., {Ichimoto}, K., \& {Suematsu}, Y. 2012, \apjl, 747, L18

\bibitem[{{Solanki} \& {Montavon}(1993)}]{1993A&A...275..283S}
{Solanki}, S.~K. \& {Montavon}, C.~A.~P. 1993, \aap, 275, 283

\bibitem[{{Thomas} {et~al.}(2002){Thomas}, {Weiss}, {Tobias}, \&
  {Brummell}}]{2002Natur.420..390T}
{Thomas}, J.~H., {Weiss}, N.~O., {Tobias}, S.~M., \& {Brummell}, N.~H. 2002,
  \nat, 420, 390

\bibitem[{{Title} {et~al.}(1993){Title}, {Frank}, {Shine}, {Tarbell}, {Topka},
  {Scharmer}, \& {Schmidt}}]{1993ApJ...403..780T}
{Title}, A.~M., {Frank}, Z.~A., {Shine}, R.~A., {et~al.} 1993, \apj, 403, 780

\bibitem[{{Tiwari} {et~al.}(2013){Tiwari}, {van Noort}, {Lagg}, \&
  {Solanki}}]{2013A&A...557A..25T}
{Tiwari}, S.~K., {van Noort}, M., {Lagg}, A., \& {Solanki}, S.~K. 2013, \aap,
  557, A25

\bibitem[{{Tsuneta} {et~al.}(2008){Tsuneta}, {Ichimoto}, {Katsukawa}, {Nagata},
  {Otsubo}, {Shimizu}, {Suematsu}, {Nakagiri}, {Noguchi}, {Tarbell}, {Title},
  {Shine}, {Rosenberg}, {Hoffmann}, {Jurcevich}, {Kushner}, {Levay}, {Lites},
  {Elmore}, {Matsushita}, {Kawaguchi}, {Saito}, {Mikami}, {Hill}, \&
  {Owens}}]{2008SoPh..249..167T}
{Tsuneta}, S., {Ichimoto}, K., {Katsukawa}, Y., {et~al.} 2008, \solphys, 249,
  167

\bibitem[{{van Noort}(2012)}]{2012A&A...548A...5V}
{van Noort}, M. 2012, \aap, 548, A5

\bibitem[{{van Noort} {et~al.}(2013){van Noort}, {Lagg}, {Tiwari}, \&
  {Solanki}}]{2013A&A...557A..24V}
{van Noort}, M., {Lagg}, A., {Tiwari}, S.~K., \& {Solanki}, S.~K. 2013, \aap,
  557, A24

\bibitem[{{Verma} {et~al.}(2018){Verma}, {Denker}, {Balthasar}, {Kuckein},
  {Rezaei}, {Sobotka}, {Deng}, {Wang}, {Tritschler}, {Collados}, {Diercke}, \&
  {Gonz{\'a}lez Manrique}}]{2018A&A...614A...2V}
{Verma}, M., {Denker}, C., {Balthasar}, H., {et~al.} 2018, \aap, 614, A2

\bibitem[{{Watanabe} {et~al.}(2014){Watanabe}, {Kitai}, \&
  {Otsuji}}]{2014ApJ...796...77W}
{Watanabe}, H., {Kitai}, R., \& {Otsuji}, K. 2014, \apj, 796, 77

\bibitem[{{Watson} {et~al.}(2014){Watson}, {Penn}, \&
  {Livingston}}]{2014ApJ...787...22W}
{Watson}, F.~T., {Penn}, M.~J., \& {Livingston}, W. 2014, \apj, 787, 22

\bibitem[{{Weiss} {et~al.}(2004){Weiss}, {Thomas}, {Brummell}, \&
  {Tobias}}]{2004ApJ...600.1073W}
{Weiss}, N.~O., {Thomas}, J.~H., {Brummell}, N.~H., \& {Tobias}, S.~M. 2004,
  \apj, 600, 1073

\bibitem[{{Wentzel}(1992)}]{1992ApJ...388..211W}
{Wentzel}, D.~G. 1992, \apj, 388, 211

\bibitem[{{Zakharov} {et~al.}(2008){Zakharov}, {Hirzberger}, {Riethm{\"u}ller},
  {Solanki}, \& {Kobel}}]{2008A&A...488L..17Z}
{Zakharov}, V., {Hirzberger}, J., {Riethm{\"u}ller}, T.~L., {Solanki}, S.~K.,
  \& {Kobel}, P. 2008, \aap, 488, L17

\end{thebibliography}

\appendix

\section{Maps of the continuum intensity of all spots} \label{sect:maps}

\begin{figure*}
\centering
\includegraphics[width=17cm]{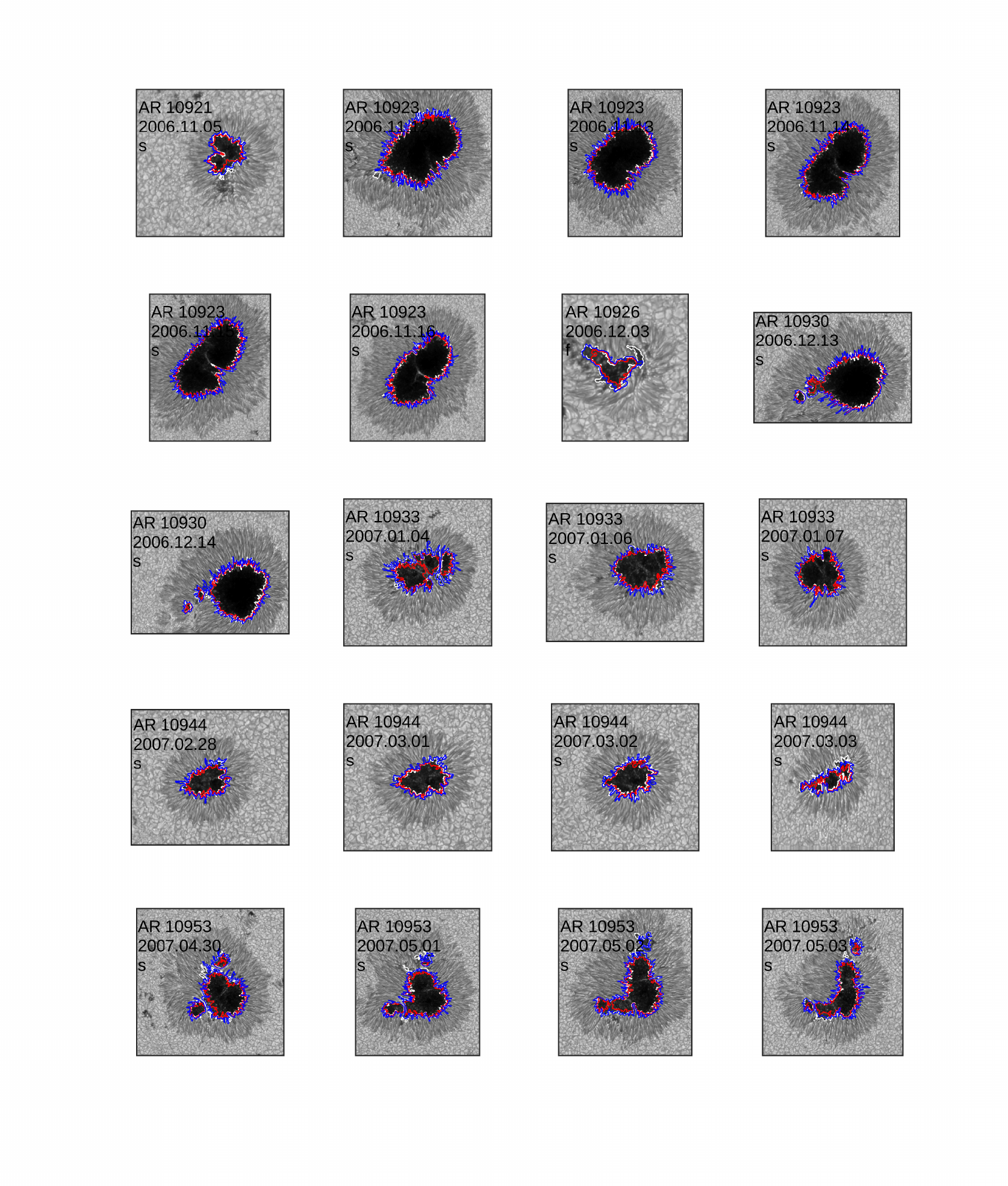}
\caption{Maps of the continuum intensity of all sunspots that were analyzed in this study. The white contours indicate a continuum intensity of 50 \% of the quiet Sun level. The two other contours correspond to different strengths of the vertical magnetic field, evaluated at $\log \tau = -0.9$. Blue: 1650~G and red: 1867~G. The letters indicate the evolutionary state of the sunspot (s: stable, f: forming, d: decaying).}
\label{fig:spots_all}
\end{figure*}

\begin{figure*}
\centering
\includegraphics[width=17cm]{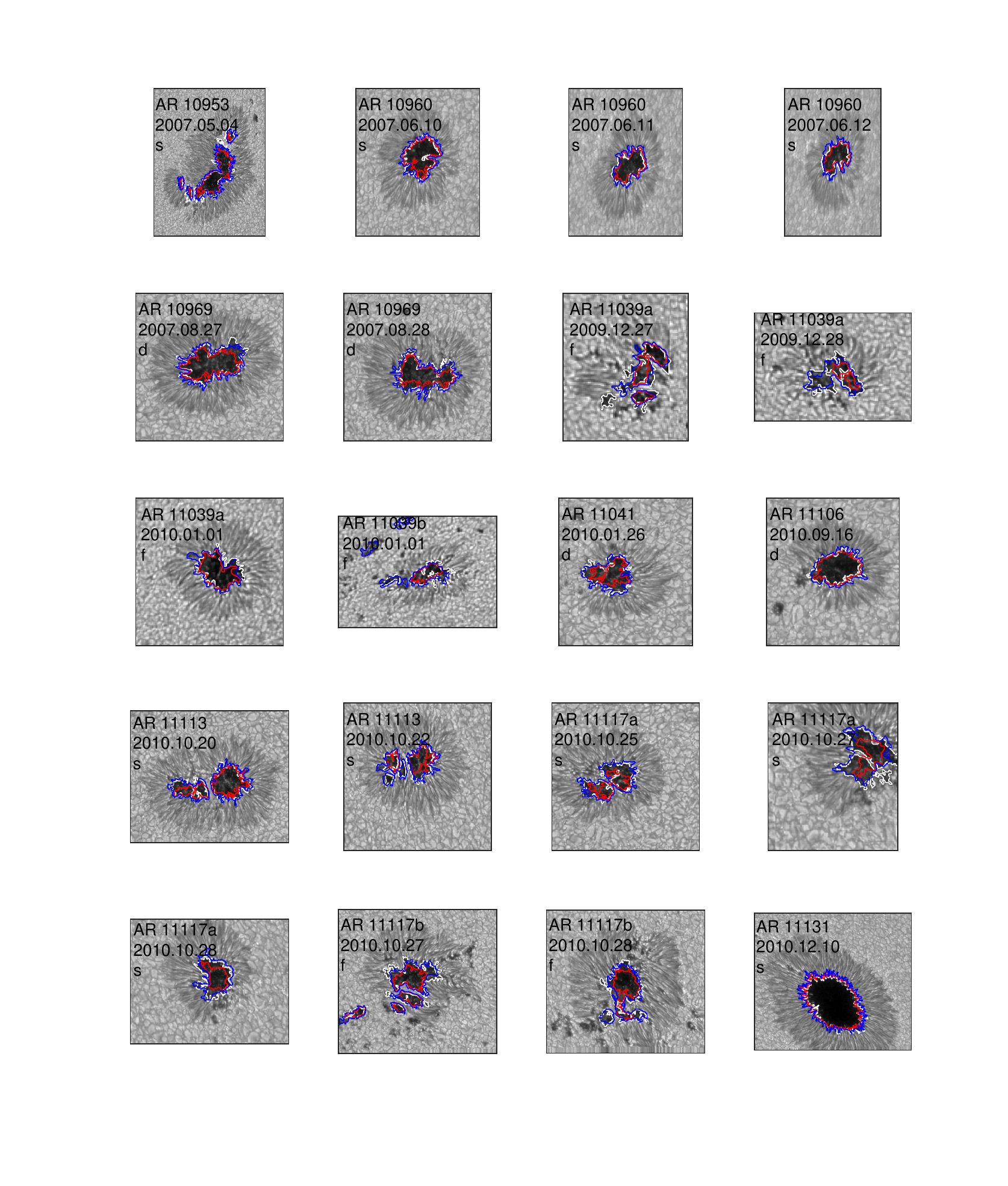}
\caption{Continuation of Figure~\ref{fig:spots_all}.}
\end{figure*}

\begin{figure*}
\centering
\includegraphics[width=17cm]{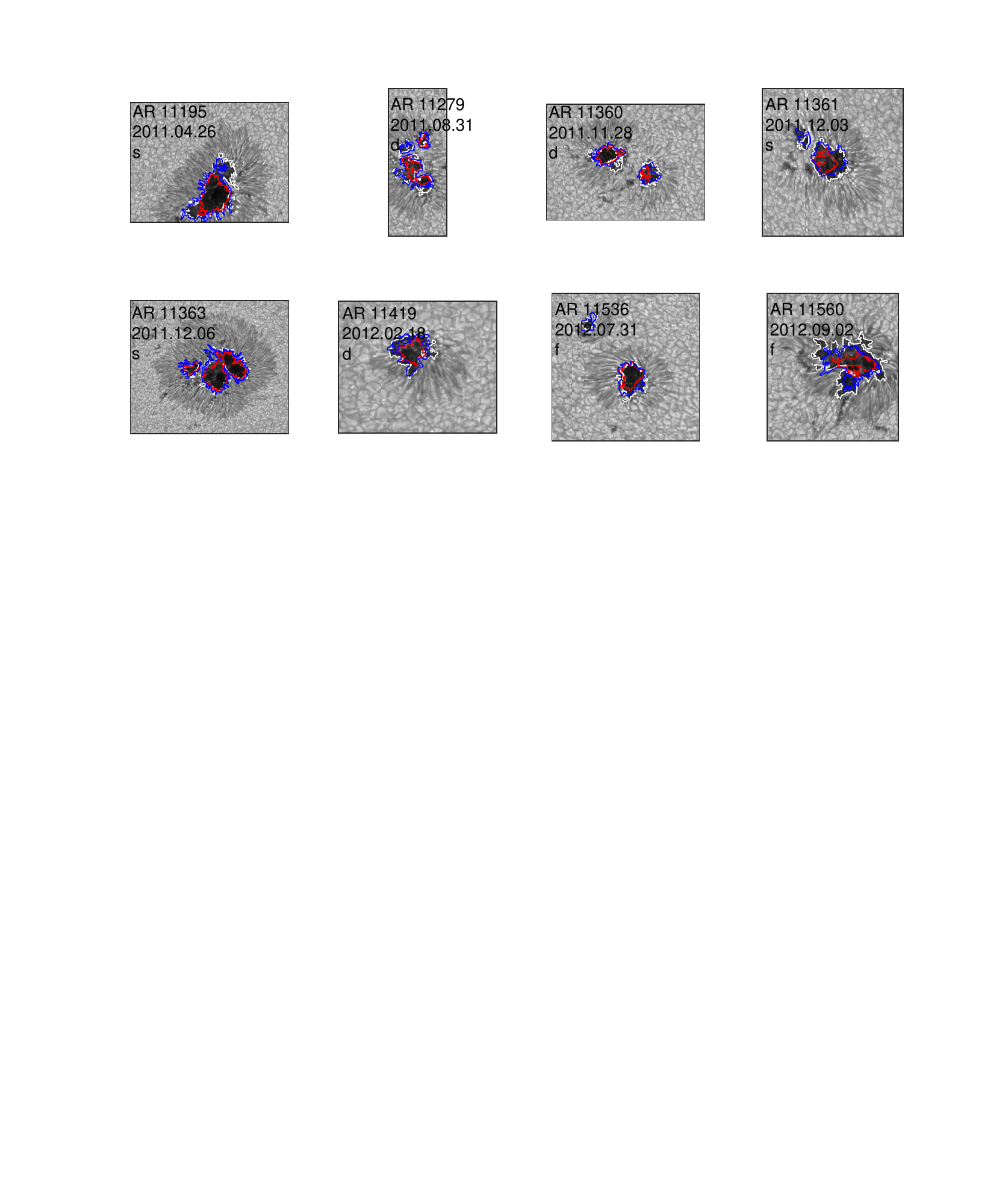}
\caption{Continuation of Figure~\ref{fig:spots_all}.}
\end{figure*}

\end{document}